\pdfoutput=1
\documentclass[aps,prd,amsmath,floats,floatfix, twocolumn,
superscriptaddress,nofootinbib,showpacs,longbibliography]{revtex4-1}
\usepackage[T1]{fontenc}
\usepackage[utf8]{inputenc}
\usepackage{txfonts}
\usepackage{verbatim}
\usepackage[dvipsnames, usenames]{xcolor}
\definecolor{linkcolor}{rgb}{0.0,0.3,0.5}
\usepackage[hypertexnames=false, unicode, colorlinks=true, linkcolor=linkcolor,
citecolor=linkcolor, filecolor=linkcolor,urlcolor=linkcolor,
pdfusetitle]{hyperref}
\usepackage[all]{hypcap}
\usepackage{graphicx}
\usepackage{xspace}
\usepackage{amssymb}
\usepackage{microtype}
\usepackage{subfigure}

\usepackage{url}

\usepackage[english]{babel}
\usepackage{blindtext}

\usepackage[normalem]{ulem} 
\usepackage{bm} 
\usepackage{mathrsfs}
\usepackage{xspace} 
\usepackage{fontawesome}

\DeclareMathAlphabet{\mathpzc}{OT1}{pzc}{m}{it}
\usepackage[nomessages]{fp}

\newcommand{\sk}[1]{}

\begin{document}
\title{Accurate models for recoil velocity distribution in black hole mergers with comparable to extreme mass-ratios and their astrophysical implications}

\author{Tousif Islam}
\email{tousifislam@ucsb.edu}
\affiliation{Kavli Institute for Theoretical Physics,University of California Santa Barbara, Kohn Hall, Lagoon Rd, Santa Barbara, CA 93106}

\author{Digvijay Wadekar}
\affiliation{Department of Physics and Astronomy, Johns Hopkins University, 3400 N. Charles Street, Baltimore, Maryland, 21218, USA}
\affiliation{\mbox{Weinberg Institute, University of Texas at Austin, Austin, TX 78712, USA}}
 
\hypersetup{pdfauthor={Islam et al.}}

\date{\today}

\begin{abstract}
Modeling the remnant recoil velocity (kick) distribution from binary black hole mergers is crucial for understanding hierarchical mergers in active galactic nuclei or globular clusters. Existing analytic models often show large discrepancies with numerical relativity (NR) data, while data-driven models are limited to mass ratios of $q \le 8$ (aligned spins) and $q \le 4$ (precessing spins) and break down when extrapolated outside their training ranges. Using $\sim 5000$ of NR simulations from the SXS and RIT catalogs up to $q=128$ and $\sim 100$ black hole perturbation theory simulations up to $q=200$, we present two classes of models: ($i$) \textcolor{linkcolor}{\texttt{gwModel\_kick\_q200}} (\textcolor{linkcolor}{\texttt{gwModel\_kick\_q200\_GPR}}), an analytic (Gaussian process regression) model for aligned-spin binaries.  ($ii$) \textcolor{linkcolor}{\texttt{gwModel\_kick\_prec\_flow}}, a normalizing-flow model for kick distribution from precessing binaries with isotropic spins. Our approach combines analytic insights from post-Newtonian theory with data-driven techniques to ensure correct limiting behavior and high accuracy across parameter space. Both \textcolor{linkcolor}{\texttt{gwModel\_kick\_q200}} and \textcolor{linkcolor}{\texttt{gwModel\_kick\_prec\_flow}} are valid from comparable to extreme mass ratios. Extensive validation shows all three models outperform existing ones within their respective domains. Finally, using both back-of-the-envelope estimates and 1404 detailed star cluster simulations incorporating our kick models, we find that the black hole retention probability in low mass globular clusters can vary noticeably when the \textcolor{linkcolor}{\texttt{gwModel\_kick\_prec\_flow}} model is employed compared to previous analytic or data-driven models.
The models are publicly available through the \textcolor{linkcolor}{\texttt{gwModels}}\footnote{\href{https://github.com/tousifislam/gwModels}{https://github.com/tousifislam/gwModels}} package.
\end{abstract}

\maketitle

\section{Introduction}
\label{sec:intro}
One of the most intriguing features of binary black hole (BBH) mergers is that the remnant black hole (BH) receives a recoil velocity—commonly referred to as a `kick'. Over the past two decades, numerous numerical and analytic studies have been conducted to investigate these kick velocities using several complementary frameworks, including numerical relativity (NR)~\cite{Baker:2006vn, Baker:2007gi, Baker:2008md,Herrmann:2006cd,Lousto:2007db,Herrmann:2007ac,Herrmann:2007ex,Herrmann:2007cwl,Holley-Bockelmann:2007hmm,Jaramillo:2011re,Koppitz:2007ev,Lousto:2008dn,Lousto:2010xk,Schnittman:2007ij,Sopuerta:2006et,Pollney:2007ss,Rezzolla:2010df,Lousto:2011kp,Lousto:2012gt,Lousto:2012su,Miller:2008en,Tichy:2007hk,Zlochower:2010sn,Healy:2014yta,Lousto:2009mf}, black hole perturbation theory (BHPT)~\cite{Nakano:2010kv,Sundararajan:2010sr,Islam:2023mob,Hughes:2004ck,Price:2013paa,Price:2011fm}, and post-Newtonian (PN)~\cite{Blanchet:2005rj,Sopuerta:2006wj,Sopuerta:2006et,Favata:2004wz,Fitchett:1983qzq,Fitchett:1984qn,Wiseman:1992dv,Kidder:1995zr} approximations. Results from NR simulations have demonstrated that the kick velocity can reach values as high as $\sim 5000~\mathrm{km/s}$ for certain spin orientations~\cite{Campanelli:2007cga,Bruegmann:2007bri, Campanelli:2007ew, Choi:2007eu, Dain:2008ck,Gonzalez:2006md,Gonzalez:2007hi,Healy:2008js}. Recently, for more finely tuned spin configurations, kick velocities as large as $\sim 30{,}000~\mathrm{km/s}$ have been reported~\cite{Healy:2022jbh}.

These recoil kicks have significant astrophysical implications. When two BHs merge, the magnitude of the kick determines whether the remnant remains to its host environment, such as a globular cluster, nuclear star cluster, or active galactic nucleus, or escapes it. If the kick velocity exceeds the escape velocity of the host, the remnant is ejected and can no longer participate in subsequent mergers within that environment~\cite{Gerosa:2016vip,Borchers:2025sid}. Conversely, if the kick is smaller, the remnant BH remains within its host and may undergo additional mergers with other BHs.
While this simplistic argument is often used in phenomenological ejection scenarios, in reality the direction of the kick also matters, as a kick directed inward toward the cluster center is less likely to lead to the ejection of the remnant BH.
Such successive or `hierarchical' mergers provide a natural pathway for the formation of increasingly massive BHs~\cite{Holley-Bockelmann:2007hmm,Berti:2012zp,Gultekin:2004pm,Gerosa:2016vip,Borchers:2025sid,Gerosa:2021hsc,Gerosa:2017kvu}. BHs formed through hierarchical mergers are expected to exhibit distinct mass and spin distributions compared to those originating from the direct collapse of massive stars~\cite{Gerosa:2021hsc,Baibhav:2020xdf,Bouffanais:2019nrw}. Interestingly, several BH candidates detected through gravitational-wave (GW) observations show properties consistent with this hierarchical formation channel~\cite{Gerosa:2020bjb}.
Furthermore, the remnant kick has already been inferred to be as large as $1542_{-1098}^{+747}~\mathrm{km/s}$ for GW200129~\cite{Varma:2022pld,Islam:2023zzj} and $485_{-252}^{+668}~\mathrm{km/s}$ for GW191109~\cite{Islam:2023zzj}. We however note that both these event suffered from coincident detector glitches which may potentially impact the kick inference~\cite{Payne:2022spz}.

Understanding the population of BHs formed through hierarchical mergers is also crucial for modeling the dynamical evolution of astrophysical systems such as globular clusters and active galactic nuclei (AGN)~\cite{Boylan-Kolchin:2004fnd,Khonji:2024apj,Blecha:2010recoil,Gerosa:2019zmo}. Such an understanding relies heavily on the availability of accurate models for the kick velocities produced in BBH mergers. Significant progress has been made in modeling kick velocities over the past decade~\cite{Zlochower:2015wga,Baker:2008md,Lousto:2008dn,Lousto:2010xk,Lousto:2012gt,Lousto:2012su,vanMeter:2010md,Healy:2014yta,Sundararajan:2010sr,Islam:2023mob,Varma:2019csw,Varma:2018aht,Merritt:2004xa,Kidder:1995zr,Sperhake:2019wwo}. Nevertheless, a substantial modeling gap remains, particularly in accurately predicting kick velocities across the full parameter space of mass ratios and spin configurations. 

\begin{figure*}
\includegraphics[width=\textwidth]{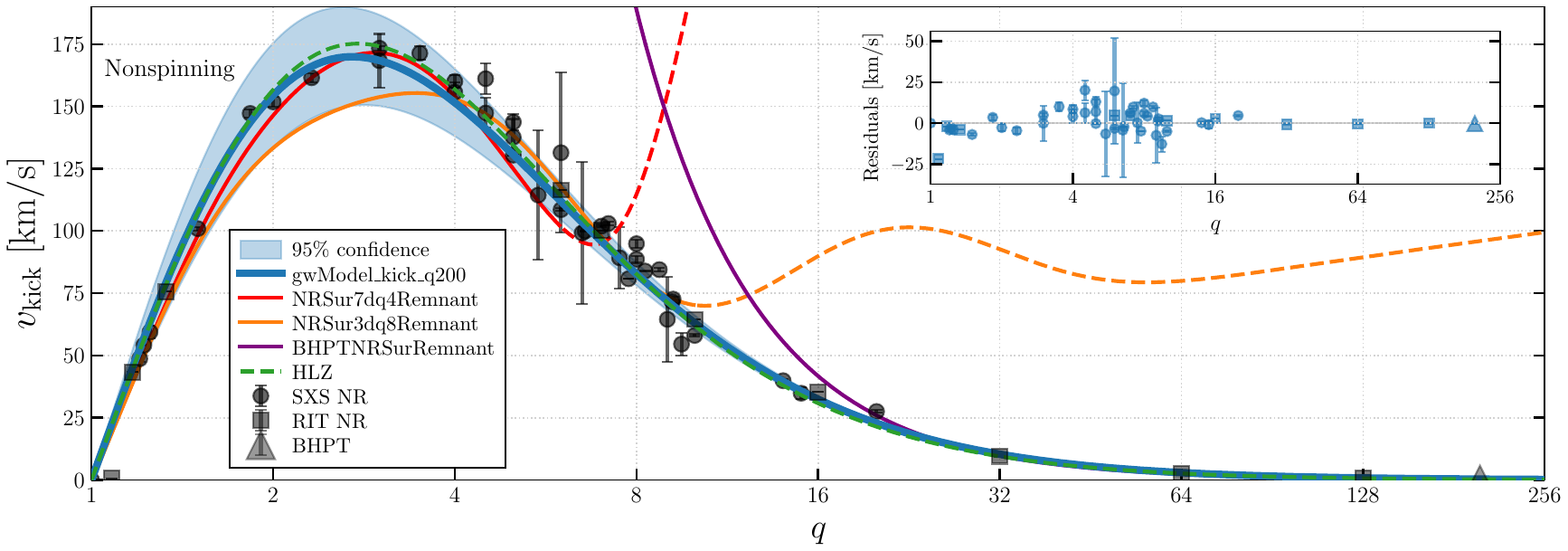}
\caption{We show the kick-velocity predictions from the analytic \textcolor{linkcolor}{\texttt{gwModel\_kick\_q200}} model (blue solid line) along with the corresponding $95\%$ confidence interval (blue shaded region) in the nonspinning limit, as a function of mass ratio from $q = 1$ to $q = 256$. For comparison, we include a combination of NR data from the SXS (circles) and RIT (squares) catalogs, as well as BHPT results (triangles). The red, orange, purple, and green lines correspond to predictions from the \textcolor{linkcolor}{\texttt{NRSur7dq4Remnant}}, \textcolor{linkcolor}{\texttt{NRSur3dq8Remnant}}, \texttt{BHPTNRSurRemnant}, and \textcolor{linkcolor}{\texttt{HLZ}} models, respectively, all shown in their nonspinning limits. The red and orange dashed lines indicate extrapolations of the \textcolor{linkcolor}{\texttt{NRSur7dq4Remnant}} and \textcolor{linkcolor}{\texttt{NRSur3dq8Remnant}} models beyond their region of validity. The inset shows the residuals of the \textcolor{linkcolor}{\texttt{gwModel\_kick\_q200}} predictions. More details are in Section~\ref{sec:aligned_spin_accuracy}.}
\label{fig:nospin}
\end{figure*}

Currently, the state-of-the-art models for kick velocities include four main frameworks. \textcolor{linkcolor}{\texttt{NRSur7dq4Remnant}}~\cite{Varma:2019csw} and \textcolor{linkcolor}{\texttt{NRSur3dq8Remnant}}~\cite{Varma:2018aht} are widely used NR surrogate models for precessing-spin and aligned-spin binaries, respectively. They are trained on NR simulations from the SXS catalog~\cite{Mroue:2013xna,Boyle:2019kee} up to mass ratios of $q = 4$ and $q = 8$ respectively, although they are often used beyond their nominal training ranges, up to $q \sim 6$ and $q \sim 10$ respectively. We define the mass ratio $q:=m_1/m_2$ where $m_1$ ($m_2$) is the mass of the larger (smaller) BH. On the other hand, \textcolor{linkcolor}{\texttt{BHPTNRSurRemnant}}~\cite{Islam:2023mob} is a remnant-kick model based on BHPT simulations, trained over mass ratios in the range $3 \leq q \leq 1000$ and calibrated against NR data, but applicable only to non-spinning binaries. All three of these models are based on Gaussian process regression (GPR) and are therefore susceptible to extrapolation errors outside their training domains. In contrast, an analytic model, commonly referred to as the \textcolor{linkcolor}{\texttt{HLZ}} model~\cite{Lousto:2008dn,Lousto:2010xk,Lousto:2012gt,Lousto:2012su,Gonzalez:2007hi}, was developed using NR data from the RIT catalog~\cite{Healy:2020vre,Healy:2022wdn}, combined with insights from PN theory and BHPT limits. While this model is, in principle, valid across a wide range of mass ratios and spin configurations, it tends to yield larger errors when compared to NR results in the comparable-mass regime. Recently, Ref.~\cite{Boschini:2023ryi} introduced remnant surrogates for the final mass and spin of BBH mergers extended to the extreme mass-ratio regime by incorporating analytical calculations in the test-particle limit, following the approach of Refs.~\cite{Hofmann:2016yih,Planas:2024vnq,Jimenez-Forteza:2016oae}. However, an analogous prescription for the recoil velocity is still lacking.

In this paper, we present new models for the kick velocity that bridge the existing gap in the parameter-space coverage and improves model accuracy. We combine approximately $5000$ NR simulations spanning mass ratios $q \in [1, 128]$ from both the SXS~\cite{Mroue:2013xna,Boyle:2019kee,Scheel:2025jct} and RIT catalogs~\cite{Healy:2020vre,Healy:2022wdn}, together with about $100$ BHPT simulations covering $q \in [40, 200]$. 
Our BHPT simulations are obtained using the time-domain Teukolsky solver developed in Refs.~\cite{Khanna:2004,Burko:2007,Sundararajan:2008zm,Sundararajan:2010sr,Ori:2000zn,Hughes:2019zmt,Apte:2019txp}. These simulations are based on linear perturbation theory, where the smaller BH undergoes an adiabatic inspiral driven by the loss of energy and angular momentum through gravitational radiation. Recent studies have demonstrated that such simulations can reproduce NR results with high accuracy in the comparable-mass regime after minimal calibration and are expected to remain reliable in the intermediate–mass-ratio range ($q \geq 15$)~\cite{Islam:2022laz,Islam:2025tjj,Islam:2022laz,Rink:2024swg,vandeMeent:2020xgc,Wardell:2021fyy,Islam:2023qyt,Islam:2023jak}.

By fusing analytic insights with data-driven techniques such as GPR and normalizing flows, we construct unified kick-velocity models that are valid from the comparable-mass to the extreme-mass-ratio regime.  Specifically, we introduce two models for aligned-spin binaries: an analytic model, \textcolor{linkcolor}{\texttt{gwModel\_kick\_q200}}, and a data-driven model, \textcolor{linkcolor}{\texttt{gwModel\_kick\_q200\_GPR}}. Both models are trained on a hybrid dataset combining NR and BHPT results across the parameter space $q \in [1, 200]$ and spin magnitudes $\chi_{1,2} \in [0, 1]$. Furthermore, we present a normalizing-flow–based probabilistic model for precessing-spin binaries, \textcolor{linkcolor}{\texttt{gwModel\_kick\_prec\_flow}}, trained over the range $q \in [1, 100]$ and $\chi_{1,2} \in [0, 1]$. Details of the aligned-spin models are provided in Section~\ref{sec:aligned_spin}, while the precessing-spin model is described in Section~\ref{sec:precessing}. We find that our models exhibit excellent extrapolation behavior and can be used even in the extreme mass ratio limit.
We discuss the implications of these new models for cluster dynamics and hierarchical BH mergers in Section~\ref{sec:implication}. 
Finally, we outline our future directions in Section~\ref{sec:conclusion}. All models presented in this work are publicly available through the \textcolor{linkcolor}{\texttt{gwModels}} package.

\section{Aligned-spin models}
\label{sec:aligned_spin}
First, we construct two kick models for aligned-spin binaries that are valid up to a mass ratio of $q = 200$ for spin magnitudes $\bold{\chi_{1,2}}\in [0,1]$. One of the models is data-driven, while the other is analytic. We refer to the analytic model as \textcolor{linkcolor}{\texttt{gwModel\_kick\_q200}}, and to the data-driven model as \textcolor{linkcolor}{\texttt{gwModel\_kick\_q200\_GPR}}, as it is trained using GPR.

Our training dataset consists of NR simulations from the SXS and RIT catalogs, together with BHPT waveforms. Specifically, we use 494 aligned-spin NR simulations from the SXS catalog with mass ratios up to $q = 20$, including 49 nonspinning cases. We also include 9 aligned-spin RIT simulations at $q = 16$ and $q = 32$, as well as 16 nonspinning RIT simulations spanning mass ratios from $q = 1.08$ to $q = 128$. In addition, we incorporate BHPT data at $q = 200$ for spin values $|\chi_1| = [0.8,\, 0.4,\, 0.0,\, -0.4,\, -0.8]$. In the nonspinning limit, multiple NR simulations are often available at the same mass ratio; in those cases, we take the average kick velocity and compute the standard deviation to estimate an uncertainty band.
These averaged data points are used only when constructing the fits. When computing the model errors, we revert to the actual kick-velocity values from the individual simulations.
For the NR data, we use the kick velocities provided directly in the corresponding catalogs, where they are computed from either horizon quantities or waveform fluxes. For the BHPT data, we compute the kick velocities using the \textcolor{linkcolor}{\texttt{gw\_remnant}}\footnote{\href{https://github.com/tousifislam/gw\_remnant}{https://github.com/tousifislam/gw\_remnant}} package~\cite{Islam:2023mob} that relies on waveform fluxes.

To develop our models, we adopt the fitting ansatz proposed in Ref.~\cite{Healy:2014yta}, which includes additional terms compared to the formulations in Refs.~\cite{Lousto:2012gt,Lousto:2012su}.
We first introduce a set of auxiliary variables.
We define the symmetric mass ratio as $\eta = q/(1+q)^2$ and the mass-difference parameter as $\delta m = (q - 1)/(q + 1)$. Furthermore, we introduce two PN-inspired auxiliary quantities, the mass-weighted effective spin parameters:
\begin{align}
\tilde{S}_{\parallel} &=
\frac{\chi_{1z} + q^2 \chi_{2z}}{(1+q)^2}, &
\Delta_{\parallel} &=
\frac{\chi_{1z} - q\, \chi_{2z}}{1+q}.
\end{align}
Additionally, we define the effective spin parameters as
\begin{align}
\chi_{\mathrm{eff}} &= \frac{q\,\chi_{1z} + \chi_{2z}}{1+q} , 
\qquad
\hat{\chi} =
\frac{\chi_{\mathrm{eff}} - 38\,\eta\,(\chi_{1z} + \chi_{2z})/113}
{1 - 76\,\eta/113} ,
\end{align}
and the anti-symmetric spin as
\begin{equation}
\chi_a = \frac{1}{2}(\chi_{1z} - \chi_{2z}) .
\tag{S6}
\end{equation}
To train the model, we use 5-fold cross-validation, dividing the data into five subsets. In each iteration, we use four subsets to train the model and the remaining held-out subset for validation.

\subsection{GPR model for aligned-spin kicks}
\label{sec:aligned_spin_gpr}
Our first model is based on gaussian process regression (GPR), as GPR-based surrogates for waveform and remnant properties have been shown to be accurate enough and are routinely used in GW data analysis~\cite{Blackman:2015pia,Varma:2018aht,Varma:2019csw,Islam:2021mha,Islam:2022laz}. We will however show in section~\ref{sec:aligned_spin_analytic} below that the analytic model instead performs better than GPR, and we therefore adopt the analytic model as the default choice in the remainder of this paper.
For the GPR model, we use the \textcolor{linkcolor}{\texttt{scikit-learn}} implementation with an optimized kernel of the form
\begin{equation}
k(\mathbf{x}, \mathbf{x}')
= 0.939^2\, k_{\mathrm{RBF}}(\mathbf{x}, \mathbf{x}')
+ k_{\mathrm{white}}(\mathbf{x}, \mathbf{x}'),
\end{equation}
where the squared exponential (RBF) component is defined as
\begin{equation}
k_{\mathrm{RBF}}(\mathbf{x}, \mathbf{x}')
= \exp\!\left[-\frac{1}{2}\sum_{i=1}^{3}
\frac{(x_i - x_i')^2}{\ell_i^2}\right],
\end{equation}
with $\boldsymbol{\ell} = [0.619,\, 2.57,\, 0.602]$. Here, $\boldsymbol{\ell}$ denotes the characteristic length scales of the input parameters $\mathbf{x} = \{\log_2(q),\, \hat{\chi},\, \chi_a\}$, which control how rapidly the correlation decays along each dimension. The amplitude term $0.939^2$ sets the overall variance of the process. The white-noise component is given by
\begin{equation}
k_{\mathrm{white}}(\mathbf{x}, \mathbf{x}')
= 0.00187\, \delta_{\mathbf{x}\mathbf{x}'},
\end{equation}
where $\delta_{\mathbf{x}\mathbf{x}'}$ is the Kronecker delta. The white-noise term accounts for uncorrelated statistical scatter or numerical noise in the training data. In this formulation, $\mathbf{x}$ and $\mathbf{x}'$ represent two points in the input feature space, and the kernel $k(\mathbf{x}, \mathbf{x}')$ quantifies their similarity, determining how strongly the corresponding outputs are correlated under the GPR model.

\begin{figure}
    \centering
    \includegraphics[width=\columnwidth]{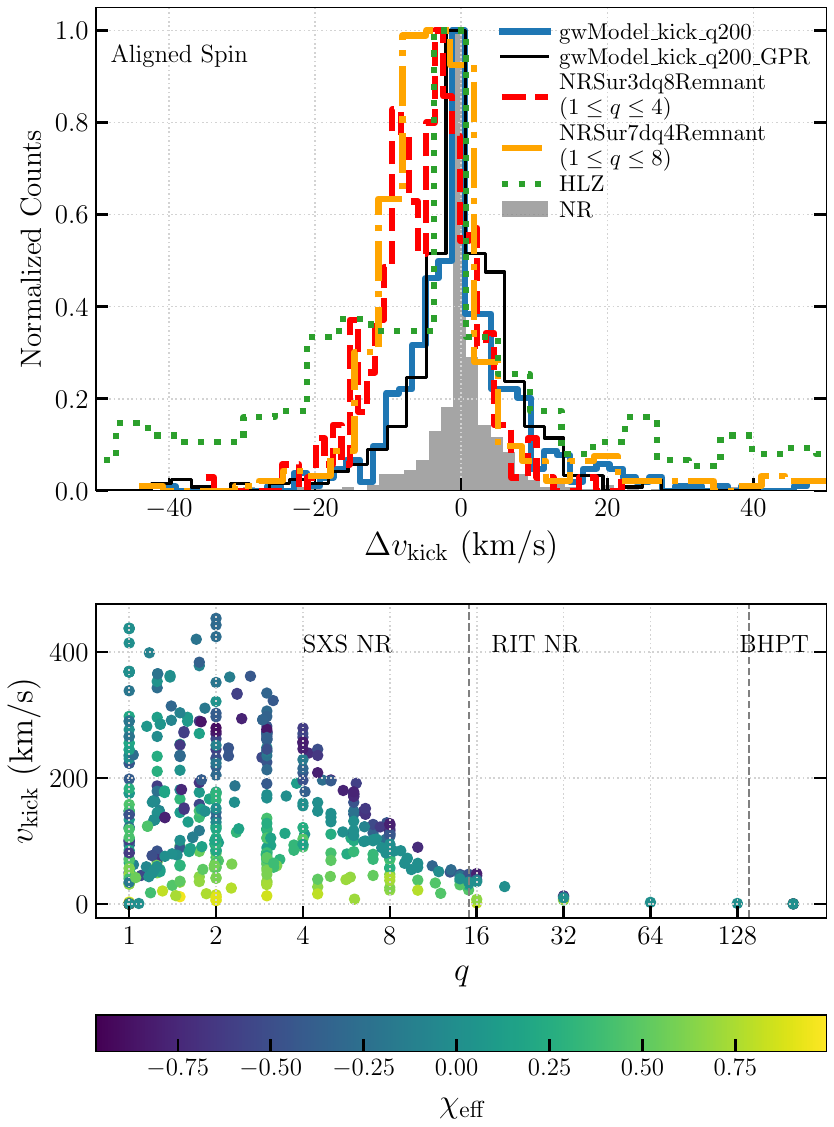}
    \caption{(Upper panel) We show the histogram of errors in kick-velocity predictions from \textcolor{linkcolor}{\texttt{gwModel\_kick\_q200}} (blue) and \textcolor{linkcolor}{\texttt{gwModel\_kick\_q200\_GPR}} (black), computed with respect to NR and BHPT results for aligned-spin binaries across five different models in their aligned-spin limit. The NR error (gray) is estimated by comparing simulations at two different numerical resolutions when available. We find that \textcolor{linkcolor}{\texttt{gwModel\_kick\_q200}} performs as well as, or in some cases better than, the NR surrogate models \textcolor{linkcolor}{\texttt{NRSur7dq4Remnant}} (orange) and \textcolor{linkcolor}{\texttt{NRSur3dq8Remnant}} (red) within their respective training domains, and shows significantly improved agreement with the data compared to the \textcolor{linkcolor}{\texttt{HLZ}} model (green). The errors from our model are comparable to the intrinsic NR uncertainties. 
    (Lower panel) We show the kick-velocity distribution as a function of the mass ratio $q$ and the effective inspiral spin $\chi_{\mathrm{eff}}$. Dashed lines indicate regions where additional data are available from SXS NR, RIT NR, and BHPT simulations. More details are in Section~\ref{sec:aligned_spin_accuracy}.}
    \label{fig:aligned_spin_accuracy}
\end{figure}

\subsection{Analytic model for aligned-spin kicks}
\label{sec:aligned_spin_analytic}
GPR models typically perform poorly when extrapolated beyond the range of the training data. To ensure reliable extrapolation and impose known physical constraints, such as the requirement of zero recoil in the equal-mass non-spinning limit, we place greater emphasis on the analytic model.
Following Refs.~\cite{Lousto:2008dn,Lousto:2010xk,Lousto:2012gt,Lousto:2012su,Gonzalez:2007hi}, we decompose the total kick velocity in aligned-spin binaries into two contributions:
\begin{equation}
V_{\rm kick,AS}(q, \chi_{1z}, \chi_{2z}) =
\sqrt{V_m^2 + V_\perp^2 + 2\, V_m V_\perp \cos\xi},
\end{equation}
where $V_m$ denotes the contribution due to mass asymmetry, $V_\perp$ is the in-plane recoil arising from spins aligned with the orbital angular momentum, and $\xi$ is the angle between the unequal-mass and spin contributions to the kick in the orbital plane.
The mass-asymmetry contribution is modeled as (following Refs.~\cite{Lousto:2008dn,Lousto:2010xk,Lousto:2012gt,Lousto:2012su,Sopuerta:2006wj})
\begin{equation}
V_m = A\,\eta^2\,\delta m\,(1 + B\,\eta + C\,\eta^2),
\label{eq:nospin_coeff}
\end{equation}
with $A = (1.3753 \pm 0.0245) \times 10^{4}\,\mathrm{km/s}$, 
$B = -\,2.636 \pm 0.168$, and $C = 5.433 \pm 0.531$.
The spin-induced contribution to the kick, arising from spins aligned with the orbital angular momentum axis, is modeled using the following polynomial expansion adopted from Ref.~\cite{Healy:2014yta}:
\begin{align}
V_{\perp} &= H\,\eta^{2}\Big(
\Delta_{\parallel}
+ H_{2a}\,\tilde{S}_{\parallel}\,\delta m
+ H_{2b}\,\Delta_{\parallel}\,\tilde{S}_{\parallel}
+ H_{3a}\,\Delta_{\parallel}^{2}\,\delta m
\nonumber\\ 
&\quad
+ H_{3b}\,\tilde{S}_{\parallel}^{2}\,\delta m
+ H_{3c}\,\Delta_{\parallel}\,\tilde{S}_{\parallel}^{2}
+ H_{3d}\,\Delta_{\parallel}^{3}
+ H_{3e}\,\Delta_{\parallel}\,\delta m^{2}
\nonumber\\
&\quad
+ H_{4a}\,\tilde{S}_{\parallel}\,\Delta_{\parallel}^{2}\,\delta m
+ H_{4b}\,\tilde{S}_{\parallel}^{3}\,\delta m
+ H_{4c}\,\tilde{S}_{\parallel}\,\delta m^{3}
\nonumber\\
&\quad
+ H_{4d}\,\Delta_{\parallel}\,\tilde{S}_{\parallel}\,\delta m^{2}
+ H_{4e}\,\Delta_{\parallel}\,\tilde{S}_{\parallel}^{3}
+ H_{4f}\,\tilde{S}_{\parallel}\,\Delta_{\parallel}^{3}
\Big),
\label{eq:vperp_poly}
\end{align}
The best-fit coefficients (averaged over 5-fold cross-validation) for the in-plane recoil, obtained using the \textcolor{linkcolor}{\texttt{scipy.curve\_fit}}~\cite{Virtanen:2020scipy} routine, are
$H = (7.402 \pm 0.028)\times10^{3}~\mathrm{km\,s^{-1}}$,
$H_{2a} = 5.891 \pm 0.042$,
$H_{2b} = -0.737 \pm 0.019$,
$H_{3a} = -0.608 \pm 0.040$,
$H_{3b} = -1.407 \pm 0.098$,
$H_{3c} = -1.462 \pm 0.078$,
$H_{3d} = (-1.67 \pm 0.66)\times10^{-2}$,
$H_{3e} = 6.768 \pm 0.080$,
$H_{4a} = -0.834 \pm 0.106$,
$H_{4b} = -2.768 \pm 0.148$,
$H_{4c} = 3.655 \pm 0.099$,
$H_{4d} = -2.099 \pm 0.145$,
$H_{4e} = 1.145 \pm 0.176$,
and $H_{4f} = 0.283 \pm 0.054$.
The angle $\xi$ is given by:
\begin{equation}
\xi = a + b\,\tilde{S}_{\parallel}
+ c\,\delta m\,\Delta_{\parallel},
\end{equation}
with $a = 146.30^{\circ} \pm 0.37^{\circ}$, 
$b = 111.92^{\circ} \pm 1.80^{\circ}$, 
and $c = 136.36^{\circ} \pm 2.94^{\circ}$.
Note that these coefficients are obtained from a 5-fold cross-
validation procedure and are significantly different from the ones used in Ref.~\cite{Healy:2014yta}. Furthermore, in Ref.~\cite{Healy:2014yta}, the coefficients entering Eq.~(\ref{eq:vperp_poly}) were fitted independently, while those appearing in Eq.~(\ref{eq:nospin_coeff}) were adopted from Refs.~\cite{Lousto:2008dn,Lousto:2010xk,Lousto:2012gt,Lousto:2012su,Sopuerta:2006wj}. In contrast, in this work, we fit all coefficients simultaneously.

\subsection{Model accuracy for aligned-spin kicks}
\label{sec:aligned_spin_accuracy}
In Figure~\ref{fig:nospin}, we show the kick-velocity predictions in the nonspinning limit from the \textcolor{linkcolor}{\texttt{gwModel\_kick\_q200}} model, along with the corresponding $95\%$ confidence interval, and compare them against a set of NR and BHPT data as a function of mass ratio ranging from $q = 1$ to $q = 256$. The $95\%$ confidence limits are derived from the covariance matrix of the fitted parameters, thereby quantifying the uncertainty in the analytical fit.
For comparison, we include predictions from four state-of-the-art models: \textcolor{linkcolor}{\texttt{NRSur7dq4Remnant}}, \textcolor{linkcolor}{\texttt{NRSur3dq8Remnant}}, \textcolor{linkcolor}{\texttt{BHPTNRSurRemnant}}, and \textcolor{linkcolor}{\texttt{HLZ}}. We find that both \textcolor{linkcolor}{\texttt{gwModel\_kick\_q200}} and \textcolor{linkcolor}{\texttt{HLZ}} reproduce the NR and BHPT data accurately across the entire mass-ratio range. The \textcolor{linkcolor}{\texttt{NRSur7dq4Remnant}} model agrees well with the data up to $q = 4$ (within its training domain); although it is sometimes used up to $q = 6$, Fig.~\ref{fig:nospin} shows that its predictions begin to deviate beyond $q = 4$ and become unreliable for $q > 6$, indicating poor extrapolation capability. Interestingly, \textcolor{linkcolor}{\texttt{NRSur3dq8Remnant}} (an aligned-spin model) fails to reproduce the NR results between $q = 1.8$ and $q = 4$, missing the well-known peak in the nonspinning kick. Furthermore, its performance degrades significantly for $q > 8$ (the upper limit of its training range). Although the model is often used up to $q = 10$, Fig.~\ref{fig:nospin} suggests that its predictions are inaccurate in the range $8 \lesssim q \lesssim 10$. Finally, \textcolor{linkcolor}{\texttt{BHPTNRSurRemnant}} fails to match the data for $q \lesssim 25$. Overall, in the nonspinning limit, \textcolor{linkcolor}{\texttt{gwModel\_kick\_q200}} provides the most accurate predictions among all models considered. The associated validation residuals, shown in the inset, remain mostly below $\sim 25\,\mathrm{km/s}$.

\begin{figure}
    \centering
    \includegraphics[width=\columnwidth]{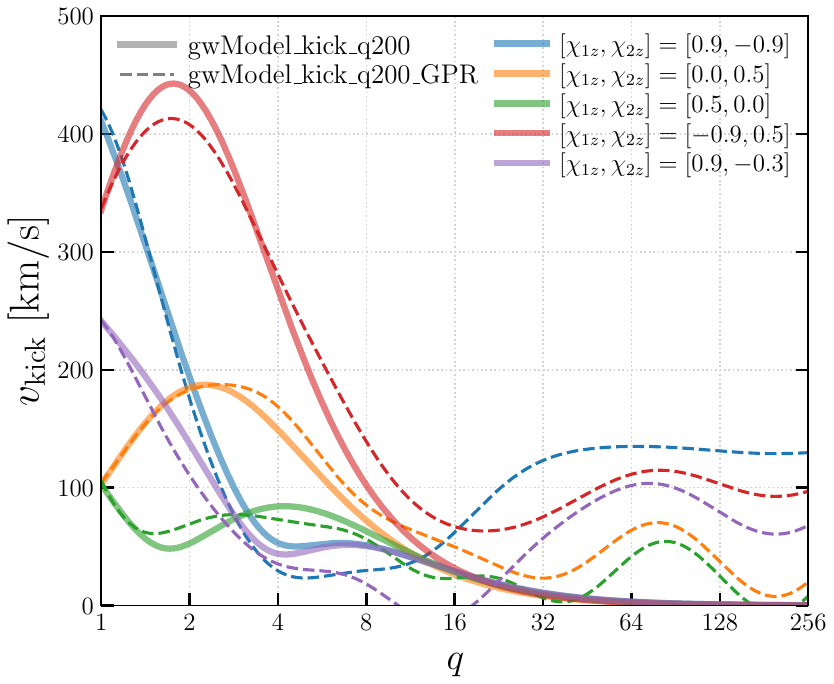}
    \caption{We show the predictions from \textcolor{linkcolor}{\texttt{gwModel\_kick\_q200}} (solid lines) and \textcolor{linkcolor}{\texttt{gwModel\_kick\_q200\_GPR}} (dashed lines) for five different aligned-spin configurations as a function of mass ratio, ranging from $q = 1$ to $q = 256$. We find that the analytic model \textcolor{linkcolor}{\texttt{gwModel\_kick\_q200}} exhibits smooth and physically consistent behavior across the full mass-ratio range, with the kick velocity decreasing monotonically with increasing $q$. In contrast, the data-driven \textcolor{linkcolor}{\texttt{gwModel\_kick\_q200\_GPR}} model produces several unphysical features, even though it achieves a slightly lower validation error. Similar unphysical trends are also observed in one-dimensional parameter-space slices for other GPR-based models, such as \textcolor{linkcolor}{\texttt{NRSur7dq4Remnant}} and \textcolor{linkcolor}{\texttt{NRSur3dq8Remnant}}. An example of such behavior for \textcolor{linkcolor}{\texttt{NRSur3dq8Remnant}} is shown in Fig.~\ref{fig:nospin}. More details are in Section~\ref{sec:aligned_spin_accuracy}.}
    \label{fig:smooth_aligned_spin_kick}
\end{figure}

Next, in Figure~\ref{fig:aligned_spin_accuracy} (upper panel), we show the model prediction errors with respect to NR and BHPT data. For \textcolor{linkcolor}{\texttt{gwModel\_kick\_q200}}, \textcolor{linkcolor}{\texttt{gwModel\_kick\_q200\_GPR}}, and \textcolor{linkcolor}{\texttt{HLZ}}, we compute the errors against all available data up to $q = 200$ (shown in the lower panel of Fig.~\ref{fig:aligned_spin_accuracy}). For \textcolor{linkcolor}{\texttt{NRSur7dq4Remnant}} and \textcolor{linkcolor}{\texttt{NRSur3dq8Remnant}}, we restrict the comparison to data points within their respective training domains, i.e., $q \leq 4$ and $q \leq 8$. We make this choice because, as shown in Fig.~\ref{fig:nospin}, the models tend to be inaccurate in the mass-ratio ranges where they are commonly extrapolated, i.e. $4 \leq q \leq 6$ and $8 \leq q \leq 10$. For reference, we also show the numerical errors in NR data, obtained by comparing kick velocities from simulations performed at different numerical resolutions whenever available. We find that both \textcolor{linkcolor}{\texttt{gwModel\_kick\_q200}} and \textcolor{linkcolor}{\texttt{gwModel\_kick\_q200\_GPR}} outperform all other models, with \textcolor{linkcolor}{\texttt{gwModel\_kick\_q200\_GPR}} showing slightly better overall accuracy. Furthermore, the errors in these two models are comparable to the intrinsic NR uncertainties. For \textcolor{linkcolor}{\texttt{gwModel\_kick\_q200}} and \textcolor{linkcolor}{\texttt{gwModel\_kick\_q200\_GPR}}, the predictions shown correspond to results from a 5-fold cross-validation procedure. The coefficient of determination ($R^2$) values for different models are as follows: $0.9958$ for NR resolution errors, $0.9919$ for \textcolor{linkcolor}{\texttt{gwModel\_kick\_q200}}, $0.9987$ for \textcolor{linkcolor}{\texttt{gwModel\_kick\_q200\_GPR}}, $0.9730$ for \textcolor{linkcolor}{\texttt{NRSur3dq8Remnant}}, $0.9948$ for \textcolor{linkcolor}{\texttt{NRSur7dq4Remnant}}, and $0.5553$ for \textcolor{linkcolor}{\texttt{HLZ}}.

Even though the \textcolor{linkcolor}{\texttt{gwModel\_kick\_q200\_GPR}} model performs slightly better than the analytic model \textcolor{linkcolor}{\texttt{gwModel\_kick\_q200}} in terms of the validation coefficient of determination, \textbf{we choose the analytic model as our default}. This choice is motivated by the fact that in the three-dimensional parameter space, several regions are sparsely populated with data. Although these regions formally lie within the training range, all models effectively extrapolate there. However, GPR-based models (such as \textcolor{linkcolor}{\texttt{gwModel\_kick\_q200\_GPR}}, \textcolor{linkcolor}{\texttt{NRSur7dq4Remnant}}, and \textcolor{linkcolor}{\texttt{NRSur3dq8Remnant}}) can often yield unreliable or nonphysical predictions in these data-sparse regions.  

We illustrate this issue in Figure~\ref{fig:smooth_aligned_spin_kick}, where we show predictions from \textcolor{linkcolor}{\texttt{gwModel\_kick\_q200}} and \textcolor{linkcolor}{\texttt{gwModel\_kick\_q200\_GPR}} for five different aligned-spin configurations, some of which are not present in the training or validation sets, across mass ratios ranging from $q = 1$ to $q = 256$. We find that predictions from the analytic model \textcolor{linkcolor}{\texttt{gwModel\_kick\_q200}} vary smoothly and decrease with increasing mass ratio, as physically expected. In contrast, the \textcolor{linkcolor}{\texttt{gwModel\_kick\_q200\_GPR}} model occasionally exhibits unphysical oscillations at large mass ratios and irregular behavior even in the comparable-mass regime. Similar artifacts are observed in the \textcolor{linkcolor}{\texttt{NRSur7dq4Remnant}} and \textcolor{linkcolor}{\texttt{NRSur3dq8Remnant}} models when inspecting one-dimensional slices of their predictions (e.g. in Fig.~\ref{fig:nospin}). These results provide a strong motivation to prefer analytic models whenever feasible.

\section{Precessing-spin model}
\label{sec:precessing}
Next, we focus on precessing-spin binaries. Modeling kicks in such systems is considerably more challenging than in aligned-spin cases, primarily because the parameter space becomes seven-dimensional owing to the four additional in-plane spin orientation angles ($\theta_1$, $\theta_2$, $\phi_1$ and $\phi_2$).
\begin{equation}
v_\mathrm{kick} = f^\mathrm{7D}(q, |\chi_1|, |\chi_2|, \theta_1,\, \theta_2,\, \phi_1, \phi_2)
\label{eq:7D_ML_model}
\end{equation}

However, if there are symmetries present in astrophysical environments, modeling the kicks can be considerably simpler. In case of black hole mergers in spherically symmetric environments such as globular clusters, the spin distribution of black holes at mergers is isotropic \cite{Gerosa:2015tea}. In such cases, we only need to model $p(v_\mathrm{kick})$, the distribution for $v_\mathrm{kick}$ marginalized over the four spin-angles, which effectively becomes a three dimensional function. We can model this using a neural network $g_\phi$ as
\begin{equation}
p(v_\mathrm{kick}) = g^\mathrm{3D}_\phi(q, |\chi_1|, |\chi_2|)
\end{equation}
where $\phi$ corresponds to the learnable weights of the neural network. Note that the 7D model in Eq.~\ref{eq:7D_ML_model} provides a point estimate, whereas the 3D case corresponds to a probability distribution; we therefore model it using normalizing flows.
We discuss the procedure to learn the weights from NR simulation data in section~\ref{sec:flow_model} below. Similarly, for disk-like environments with azimuthal symmetry, one would need to model a five dimensional function marginalized over the azimuthal angles
\begin{equation}
p(v_\mathrm{kick}) = g^\mathrm{5D}_\phi(q, |\chi_1|, |\chi_2|, \theta_1, \theta_2)
\end{equation}
In this paper, we only focus on modeling the 3D function, and we leave the 5D or 7D models to a future paper. The 3D case is useful in various astrophysical applications where there is isotropic symmetry, such as formation of compact object binaries in globular or nuclear star clusters, mergers of supermassive black holes after galaxy mergers, and early-Universe cosmological studies, where a single point estimate of the kick velocity for a specific precessing configuration is not typically required, but spin-angle–averaged recoil distributions are needed.

We build the 3D model employing a probabilistic approach based on normalizing flows. 
Our training dataset consists of a total of 2,866 SXS NR simulations covering mass ratios $q \in [1, 15]$, 1,881 RIT NR simulations over the same range, and 400 BHPT simulations spanning $q \in [40, 100]$. These datasets collectively sample a wide variety of spin configurations characterized by the spin orientation angles $(\theta_1,\, \theta_2,\, \phi_1,\, \phi_2)$ and spin magnitudes $(|\chi_1|,\, |\chi_2|)$. Interestingly, most of the NR data from the SXS and RIT catalogs are concentrated at relatively high spin magnitudes, $\chi_{1,2} \gtrsim 0.7$. 
By examining all available data, we first constrain an empirical upper bound on the kick velocities (for various spin magnitudes and orientations) as a function of the mass ratio. We express this relation as
\begin{equation}
V_{\mathrm{upper}} = V_{\mathrm{max}}\, \eta^2,
\label{eq:vupper}
\end{equation}
where $V_{\mathrm{max}} = 7.6 \times 10^{4}\,\mathrm{km/s}$. This expression is inspired by the leading-order PN behavior described in Eq.~(\ref{eq:nospin_coeff}). We emphasize that this relation serves only as a simple heuristic for understanding the envelope of kick velocities and is not used directly in our analysis.

\begin{figure}
    \centering
    \includegraphics[width=\columnwidth]{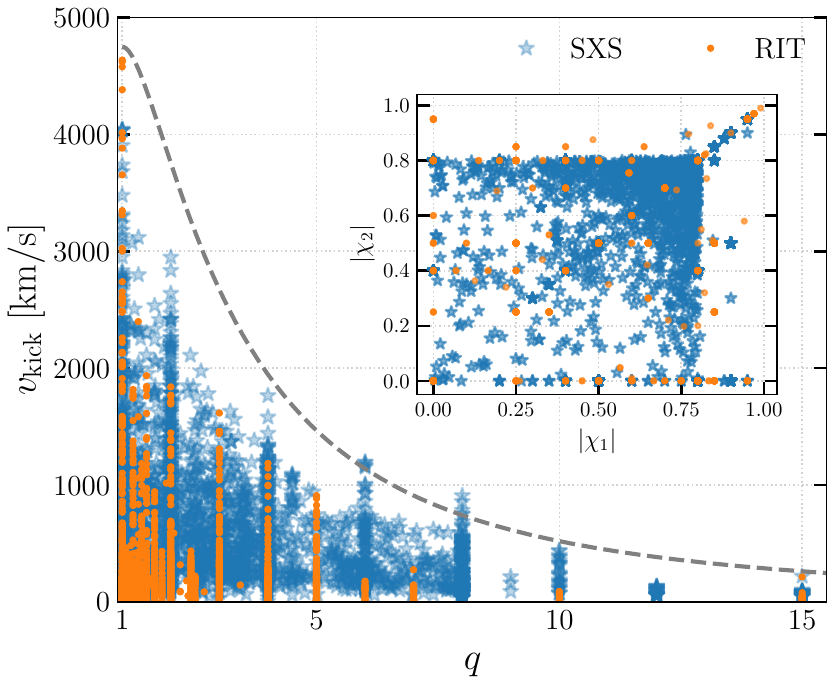}
    \caption{We show the kick-velocity distribution from precessing-spin SXS (blue stars) and RIT (orange circles) NR simulations up to $q = 15$ for different spin magnitudes and orientation configurations. Our training dataset also includes BHPT simulations spanning mass ratios from $q = 40$ to $q = 100$ (not shown). For reference, we plot the empirical upper envelope of the kick velocities as a dashed gray line, given by $V_{\mathrm{upper}} = 7.6 \times 10^{4} \eta^{2}$. The inset shows the distribution of these data points as a function of the spin magnitudes $|\chi_1|$ and $|\chi_2|$. More details are in Section~\ref{sec:flow_model_accuracy}.}
    \label{fig:precessing_spin_kick}
\end{figure}

\subsection{Normalizing flow model for precessing-spin kicks}
\label{sec:flow_model}
To model the distribution of kick velocities for precessing-spin binaries, we employ a probabilistic normalizing-flow architecture implemented using the \textcolor{linkcolor}{\texttt{nflows}} package~\cite{durkan2019neuralsplineflows}. The goal of this model is to learn the conditional probability density
$p(v_{\mathrm{kick}}\,|\,\mathbf{c})$,
where $v_{\mathrm{kick}}$ denotes the kick velocity (the target variable), and $\mathbf{c}$ represents the contextual parameters $\mathbf{c} = \{\log_2(q),\, |\chi_1|,\, |\chi_2|\}$ encoding the binary’s mass ratio and individual spin magnitudes. For any specified $(q,\, |\chi_1|,\, |\chi_2|)$, the model predicts the full probability distribution of kick velocities marginalized over isotropically-oriented spin-angle configurations. This conditional setup enables the model to efficiently capture the statistical effects of spin precession on the kick distribution while remaining computationally tractable.

\begin{figure}
    \centering
    \includegraphics[width=\columnwidth]{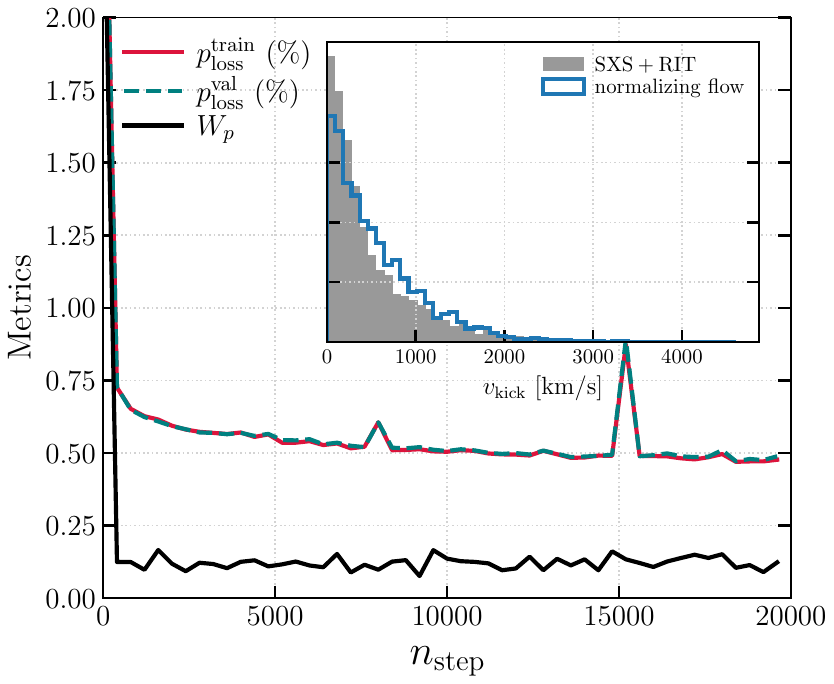}
    \caption{We show the training ($p_{\rm loss}^{\rm train}$; crimson solid line) and validation ($p_{\rm loss}^{\rm val}$; green dashed line) losses for the precessing-spin normalizing-flow model \textcolor{linkcolor}{\texttt{gwModel\_kick\_prec\_flow}} as a function of training steps. Additionally, we plot the Wasserstein distance ($W_p$; black solid line) between the precessing-spin training data and the corresponding \textcolor{linkcolor}{\texttt{gwModel\_kick\_prec\_flow}} outputs at various training stages. The inset shows the distribution of these data points as a histogram (gray), along with the corresponding distribution predicted by the trained normalizing-flow model \textcolor{linkcolor}{\texttt{gwModel\_kick\_prec\_flow}} (blue). More details are in Section~\ref{sec:flow_model_accuracy}.}
    \label{fig:nflow_loss}
\end{figure}

Before training, all quantities are standardized to zero mean and unit variance. To avoid discontinuities near the equal-mass boundary ($q=1$), we duplicate the training samples with transformations $(\log_2 q \rightarrow -\log_2 q,\; |\chi_1| \leftrightarrow |\chi_2|)$. This symmetrization ensures that the normalizing flow can learn the distribution smoothly across the parameter space. The resulting dataset is randomly shuffled and split into training ($75\%$) and validation ($25\%$) subsets.

The flow is defined as a composition of $n_{\mathrm{layers}} = 2$ autoregressive transformations, each consisting of a Masked Affine Autoregressive Transform (MAF) preceded by a \textcolor{linkcolor}{\texttt{ReversePermutation}} layer to enhance expressivity:
\begin{equation}
f = f_2 \circ f_1 = 
\big(\mathcal{P}_2 \circ \mathcal{T}_2\big) \circ 
\big(\mathcal{P}_1 \circ \mathcal{T}_1\big),
\end{equation}
where $\mathcal{P}_i$ denotes a permutation of the input features and $\mathcal{T}_i$ represents the autoregressive affine transformation. The overall flow thus defines an invertible mapping between a base latent variable $\mathbf{z}$ and the data variable $v_{\mathrm{kick}}$:
\begin{equation}
v_{\mathrm{kick}} = f(\mathbf{z};\,\mathbf{c}), \qquad \mathbf{z} \sim \mathcal{N}(0, I).
\end{equation}
The conditional likelihood is then given by
\begin{equation}
\log p(v_{\mathrm{kick}}|\mathbf{c})
= \log p_{\mathcal{Z}}\!\big(f^{-1}(v_{\mathrm{kick}};\mathbf{c})\big)
+ \log\!\left|\det\!\frac{\partial f^{-1}}{\partial v_{\mathrm{kick}}}\right|.
\end{equation}

\begin{figure}
\centering
\includegraphics[width=\columnwidth]{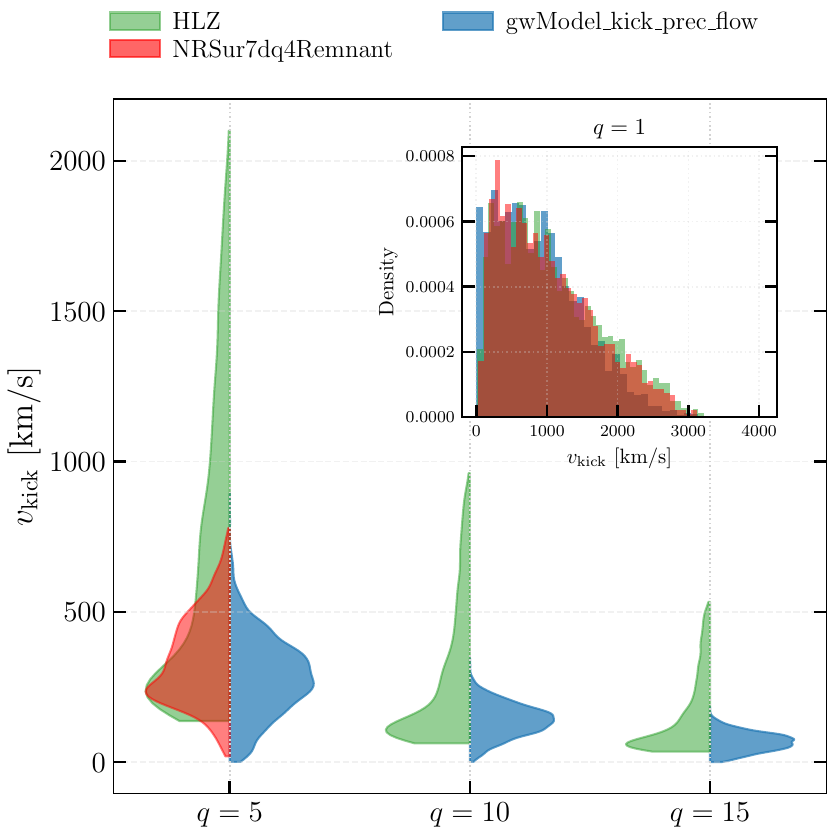}
\caption{We show the distribution of kick velocities predicted by three precessing-spin models—\textcolor{linkcolor}{\texttt{gwModel\_kick\_prec\_flow}} (blue), \textcolor{linkcolor}{\texttt{NRSur7dq4Remnant}} (red), and \textcolor{linkcolor}{\texttt{HLZ}} (green)—within their respective domains of validity, for four different mass ratios: $q = [1,\,5,\,10,\,15]$ (with the $q = 1$ results shown in the inset), assuming spin magnitudes of $[|\chi_1|,\,|\chi_2|] = [0.6,\,0.9]$. More details are in Section~\ref{sec:flow_model_accuracy}.}
\label{fig:gwModel_prec_flow_distribution}
\end{figure}

Each autoregressive block employs a fully connected multilayer perceptron (MLP) with $d_{\mathrm{hidden}} = 8$ neurons per layer and Gaussian Error Linear Unit (GELU) activation functions. The context features $\mathbf{c}$ are embedded through linear conditioning networks with dimensionality $d_{\mathrm{context}} = 3$. The base distribution is a standard normal, $p_{\mathcal{Z}} = \mathcal{N}(0, 1)$.

The model is trained by minimizing the negative log-likelihood over the training set:
\begin{equation}
\mathcal{L} = -\,\mathbb{E}_{(v_{\mathrm{kick}},\,\mathbf{c})}
\big[\log p(v_{\mathrm{kick}}|\mathbf{c})\big],
\end{equation}
where $\mathbb{E}$ is the expectation operator that denotes an average over the empirical training distribution of the data. In practice, this expectation is approximated by a Monte Carlo estimate over the finite training set,
\begin{equation}
\mathcal{L}
\approx
-\frac{1}{N}\sum_{i=1}^{N}
\log p\!\left(v_{\mathrm{kick}}^{(i)} \mid \mathbf{c}^{(i)}\right),
\end{equation}
where $\{(v_{\mathrm{kick}}^{(i)}, \mathbf{c}^{(i)})\}_{i=1}^{N}$ are the
training samples. Basically, the loss (given by the negative log likelihood) will be higher if the probability distribution predicted by the model $p$ is different from the true data distribution. During optimization, the sum is evaluated over mini-batches of size $B$,
yielding the stochastic loss
\begin{equation}
\mathcal{L}_{\mathrm{batch}}
=
-\frac{1}{B}\sum_{j=1}^{B}
\log p\!\left(v_{\mathrm{kick}}^{(j)} \mid \mathbf{c}^{(j)}\right).
\end{equation}
This stochastic estimate provides an unbiased approximation to the full negative log-likelihood objective.
We use the Adam optimizer with a learning rate of $10^{-3}$ and a weight decay of $10^{-4}$ for regularization. Training is performed for $20{,}000$ iterations with a batch size of $128$. During training, we periodically evaluate the validation loss and compute the Wasserstein distance between the generated and target kick distributions to monitor convergence and sample quality. The best model checkpoint is selected based on the minimum validation loss.
Note that we keep the flow model intentionally simple, employing only two layers. This design choice helps prevent overfitting and allows the model to learn the underlying physical manifold of the kick-velocity distribution more effectively.

\begin{figure*}
\includegraphics[width=0.66\textwidth]{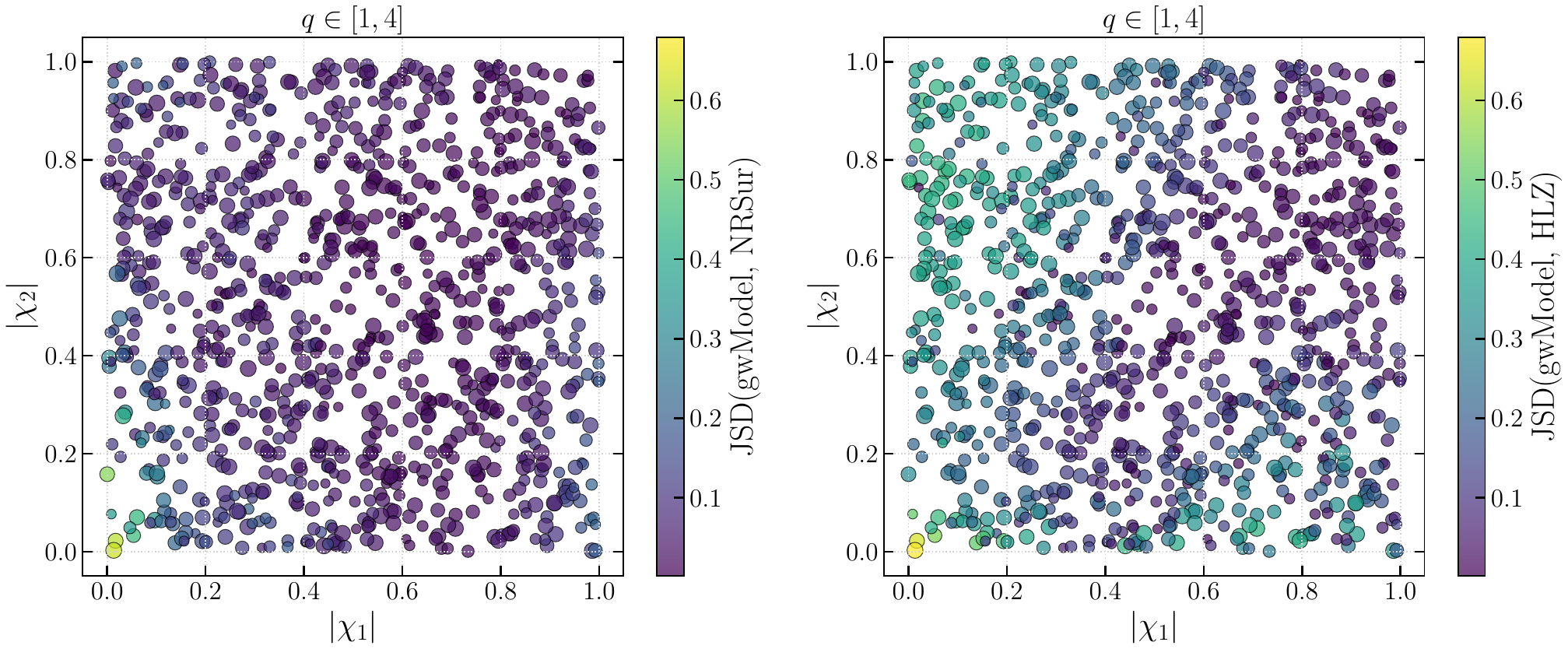}
\includegraphics[width=0.33\textwidth]{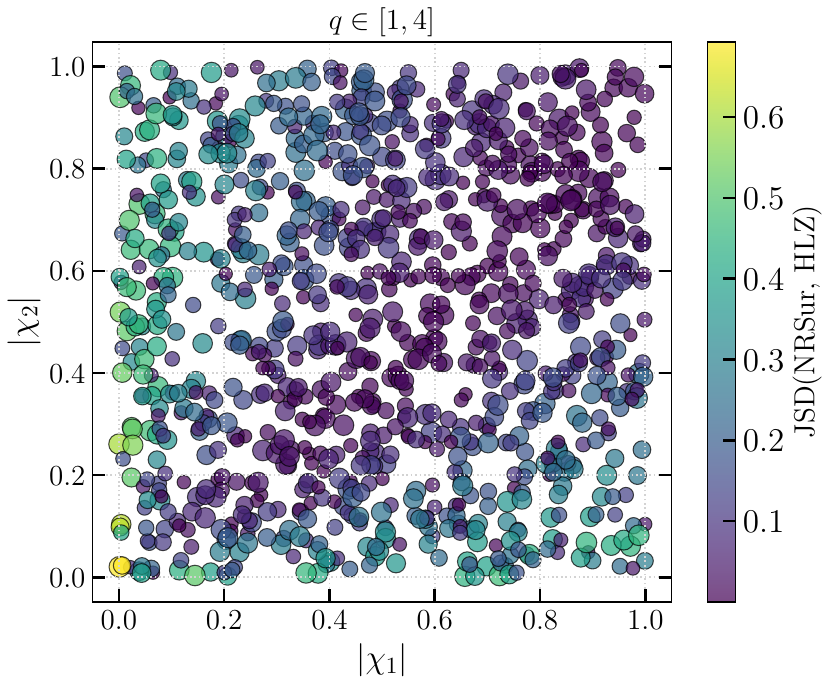}
\caption{We show the Jensen–Shannon divergence (JSD; color-coded) between the kick-velocity distributions obtained for $1000$ randomly chosen combinations of spin magnitudes $|\chi_1|$ and $|\chi_2|$. Spin orientation angles are drawn from an isotropic distribution. The left and center panels show the results for mass ratios $q$ within the range $q \in [1, 4]$, where all three models are valid: \textcolor{linkcolor}{\texttt{gwModel\_kick\_prec\_flow}} (referred to as \texttt{gwModel}), \textcolor{linkcolor}{\texttt{NRSur7dq4Remnant}} (referred to as \texttt{NRSur}), and \textcolor{linkcolor}{\texttt{HLZ}}. The symbol size scales with the mass ratio: smaller circles correspond to smaller $q$, while larger circles correspond to larger $q$. The predictions of \texttt{gwModel} and \texttt{NRSur} agree more closely compared to those of \texttt{HLZ}. More details are in Section~\ref{sec:flow_model_accuracy}.}
\label{fig:jsd}
\end{figure*}

The distribution of the spin angles $\theta_{1,2}$ and $\phi_{1,2}$ becomes the implicit prior distribution $p_{\mathrm{train}}(\mathbf{c})$ for the flow model. We however find that this prior is very close to the isotropic spin angle distribution where $\cos \theta_{1,2} \propto \mathcal{U}(0,1)$ and $\phi_{1,2} \propto \mathcal{U}(0,1)$. Because a conditional normalising flow learns the likelihood
$p(v_{\mathrm{kick}} \mid \mathbf{c})$ under the training distribution, the model predictions are calibrated with respect to this prior. If one requires the distribution of kick velocities under a different astrophysical prior $p_{\mathrm{target}}(\mathbf{c})$, one can utilize standard importance reweighting.

The final model, denoted as \textcolor{linkcolor}{\texttt{gwModel\_kick\_prec\_flow}}, captures the conditional distribution of kick velocities across a wide range of mass ratios ($q \leq 200$) and spin magnitudes. 

\subsection{Model accuracy for precessing-spin kicks}
\label{sec:flow_model_accuracy}
In Figure~\ref{fig:precessing_spin_kick}, we show the kick-velocity distribution from precessing-spin SXS and RIT NR simulations up to $q = 15$, covering a wide range of spin magnitudes and orientation configurations. We also plot the empirical upper envelope of the kick velocities, given by Eq.~(\ref{eq:vupper}). 

\begin{figure}
\centering
\includegraphics[width=\columnwidth]{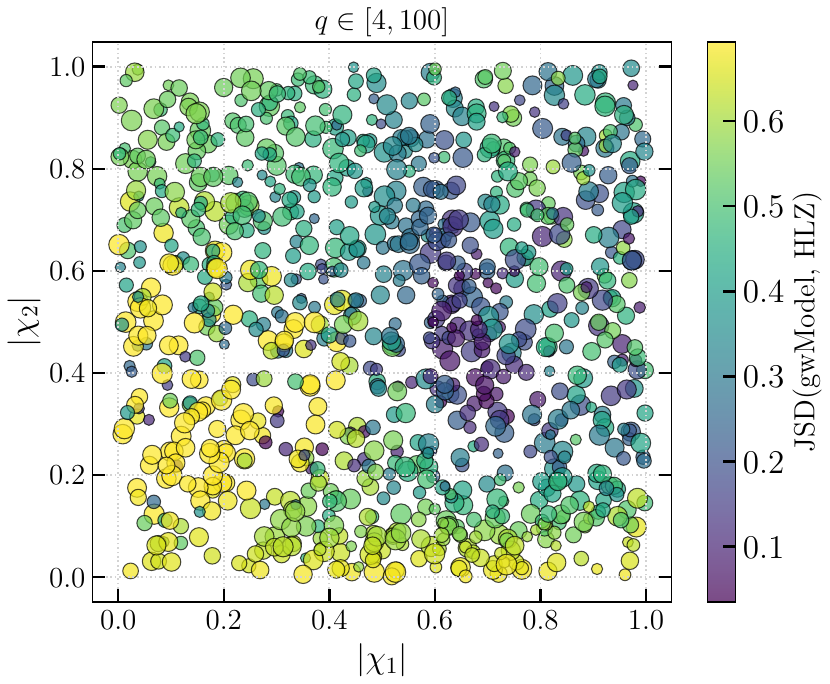}
\caption{Same as Fig.~\ref{fig:jsd} but for more asymmetric mass ratios: $q \in [4, 100]$. We do not compare with \textcolor{linkcolor}{\texttt{NRSur7dq4Remnant}} as this range is beyond its domain of validity. Spin orientation angles are again drawn from an isotropic distribution. The symbol size scales with the mass ratio: smaller circles correspond to smaller $q$, while larger circles correspond to larger $q$. More details are in Section~\ref{sec:flow_model_accuracy}.}
\label{fig:jsd_precessing_models_largeq}
\end{figure}

In Figure~\ref{fig:nflow_loss}, we present the training and validation losses for \textcolor{linkcolor}{\texttt{gwModel\_kick\_prec\_flow}} as a function of training iterations. Both losses decrease steadily and reach a value of approximately $0.5\%$ after about $5000$ training steps, indicating that the model has converged efficiently. The close agreement between the training and validation losses demonstrates that the model does not suffer from overfitting. In addition, we compute the Wasserstein distance between the precessing-spin training data and the corresponding \textcolor{linkcolor}{\texttt{gwModel\_kick\_prec\_flow}} samples at different stages of training. This distance stabilizes after roughly $1000$ iterations at a value of $\sim 0.1$, further confirming that the trained model accurately reproduces the underlying data distribution. Furthermore, we display the distribution of the kick velocities as a histogram (in the inset), along with the corresponding distribution predicted by the trained normalizing-flow model \textcolor{linkcolor}{\texttt{gwModel\_kick\_prec\_flow}}. We find that the predictions from \textcolor{linkcolor}{\texttt{gwModel\_kick\_prec\_flow}} closely reproduce the training data across the entire parameter space.

Next, in Figure~\ref{fig:gwModel_prec_flow_distribution}, we show the distribution of kick velocities predicted by three precessing-spin models—\textcolor{linkcolor}{\texttt{gwModel\_kick\_prec\_flow}}, \textcolor{linkcolor}{\texttt{NRSur7dq4Remnant}}, and \textcolor{linkcolor}{\texttt{HLZ}}—within their respective domains of validity, for four different mass ratios: $q = [1,\, 5,\, 10,\, 15]$. We fix the spin magnitudes to $[\chi_1,\, |\chi_2|] = [0.6,\, 0.9]$ and draw the spin orientation angles from an isotropic distribution. Because \textcolor{linkcolor}{\texttt{gwModel\_kick\_prec\_flow}} was trained using a different spin–orientation prior, we reweight its predictions to match an isotropic spin–orientation distribution.
For $q = 1$, all three models produce nearly identical kick–velocity distributions. At $q = 5$, \textcolor{linkcolor}{\texttt{gwModel\_kick\_prec\_flow}} and \textcolor{linkcolor}{\texttt{NRSur7dq4Remnant}} yield consistent results, whereas \textcolor{linkcolor}{\texttt{HLZ}} predicts a noticeably broader distribution, corresponding to larger recoil velocities. For higher mass ratios ($q = 10$ and $q = 15$), \textcolor{linkcolor}{\texttt{NRSur7dq4Remnant}} fails to provide meaningful predictions as these values lie beyond its training range, while the \textcolor{linkcolor}{\texttt{HLZ}} model continues to overestimate the width of the distribution compared to \textcolor{linkcolor}{\texttt{gwModel\_kick\_prec\_flow}}. 
We also observe that for lower spin magnitudes, the discrepancies between model predictions become more pronounced. This behavior likely arises because the available NR training data in this regime are sparse, causing the models to extrapolate beyond well-sampled regions of the parameter space.

We explore this comparison further in Figure~\ref{fig:jsd} (first two panels). We sample $1000$ different mass ratios within the comparable-mass regime, specifically for $q \in [1, 4]$, and spin magnitudes $|\chi_1|$ and $|\chi_2|$ uniformly distributed between $0$ and $1$. For each configuration, we compute the kick-velocity distributions obtained from isotropically distributed spin-orientation angles using three precessing-spin models: \textcolor{linkcolor}{\texttt{gwModel\_kick\_prec\_flow}}, \textcolor{linkcolor}{\texttt{NRSur7dq4Remnant}}, and \textcolor{linkcolor}{\texttt{HLZ}}.  
To quantify the similarity between the kick-velocity distributions predicted by these models, we compute the Jensen–Shannon divergence (JSD). The JSD between two probability distributions $p_1$ and $p_2$ is defined as  
\begin{equation}
\mathrm{JSD}(p_1||p_2) = \tfrac{1}{2}\mathrm{KL}(p_1||m) + \tfrac{1}{2}\mathrm{KL}(p_2||m),
\end{equation}
where $m = \tfrac{1}{2}(p_1 + p_2)$ is the pointwise mean of $p_1$ and $p_2$ and $\mathrm{KL}$ denotes the Kullback–Leibler divergence. The JSD is symmetric and bounded as $0 \leq \mathrm{JS}(P||Q) \leq \log 2 \approx 0.693$. Typically, $\mathrm{JS} \lesssim 0.1$ indicates that the two distributions are statistically similar, while values $\mathrm{JS} \gtrsim 0.1$ signify significant discrepancies between them.

We find that, across most of the parameter space, the kick-velocity distributions predicted by \textcolor{linkcolor}{\texttt{gwModel\_kick\_prec\_flow}} and \textcolor{linkcolor}{\texttt{NRSur7dq4Remnant}} agree well, with JSD$\leq 0.1$ for the majority of sampled points.

However, we see that for smaller spin magnitudes, the distributions from these two models begin to deviate. This behavior arises partly because most of the available NR training data correspond to high spin magnitudes, making all models effectively extrapolate in the low-spin regime. 
In general, there is a trade-off between ML models. The larger the ML model, the more chance it has of overfitting in the regions when data is sparse. However, on the other side, larger models can fit complex features better when sufficient training data is present. We tried to balance these cases in our flow model by including only two layers (see section~\ref{sec:flow_model}). 
As shown earlier, the more complex GPR-based models tend to perform poorly in data-sparse regions, where as the flow model is also not completely immune to data sparsity. Having more NR simulations in the low spin regime will help in building robust models. When comparing \textcolor{linkcolor}{\texttt{gwModel\_kick\_prec\_flow}} with \textcolor{linkcolor}{\texttt{HLZ}}, we find that the JSD values are significantly larger than in the previous case, indicating that the kick-velocity distributions predicted by these two models differ substantially across most of the parameter space. This comparison further suggests that \textcolor{linkcolor}{\texttt{gwModel\_kick\_prec\_flow}} performs accurately in the comparable-mass regime and agrees closely with \textcolor{linkcolor}{\texttt{NRSur7dq4Remnant}} within this range. The advantage of \textcolor{linkcolor}{\texttt{gwModel\_kick\_prec\_flow}} is that it extends to extreme mass ratios which \textcolor{linkcolor}{\texttt{NRSur7dq4Remnant}} does not. We also compare the distributions obtained from the \textcolor{linkcolor}{\texttt{NRSur7dq4Remnant}} and \textcolor{linkcolor}{\texttt{HLZ}} models and find that they differ in the regime where one of the spin magnitudes is small.

We repeat the same experiment for large mass ratios ($q \in [4,\,100]$), computing the Jensen–Shannon divergence between the kick-velocity distributions predicted by \textcolor{linkcolor}{\texttt{gwModel\_kick\_prec\_flow}} and \textcolor{linkcolor}{\texttt{HLZ}} (Figure~\ref{fig:jsd_precessing_models_largeq}). The distributions are obtained assuming isotropic spin orientations and spin magnitudes $|\chi_1|$ and $|\chi_2|$ uniformly sampled within $[0,\,1]$. We find that, across most of the parameter space, the kick-velocity distributions from these two models differ substantially, yielding JSD values significantly greater than $0.1$.

\begin{table}[h]
\centering
\caption{Median evaluation cost and standard deviation (in seconds) for computing kick-velocity distributions using a single CPU for 50 randomly sampled mass ratios within $q \in [1, 4]$ and spin magnitudes $|\chi_{1,2}| \in [-1, 1]$, with (2500) spin-orientation angles drawn from an isotropic distribution for three different models.}
\begin{tabular}{lcc}
\hline\hline
Model & Median [s] & Std. Dev. [s] \\
\hline
\textcolor{linkcolor}{\texttt{gwModel\_kick\_prec\_flow}}       & 0.0488 & 0.0016 \\
\textcolor{linkcolor}{\texttt{HLZ}}                     & 0.00024 & 0.00001 \\
\textcolor{linkcolor}{\texttt{NRSur7dq4Remnant}}          & 2.7281 & 0.1574 \\
\hline\hline
\end{tabular}
\label{tab:evaluation_cost_stats}
\end{table}

\subsection{Computational speedup}
\label{sec:speedup}
We now evaluate the computational efficiency of our precessing-spin model \textcolor{linkcolor}{\texttt{gwModel\_kick\_prec\_flow}} and compare it with two existing models: \textcolor{linkcolor}{\texttt{NRSur7dq4Remnant}} and \textcolor{linkcolor}{\texttt{HLZ}}. While \textcolor{linkcolor}{\texttt{gwModel\_kick\_prec\_flow}} is a normalizing-flow–based model, \textcolor{linkcolor}{\texttt{NRSur7dq4Remnant}} relies on GPR, and \textcolor{linkcolor}{\texttt{HLZ}} is fully analytic.  

On a single CPU, we randomly sample 50 mass ratios within $q \in [1, 4]$ and spin magnitudes $|\chi_{1,2}| \in [-1, 1]$. For each configuration, we compute the kick-velocity distribution resulting from 2,500 isotropically distributed spin orientations. We use 2,500 points to ensure that our timing results have converged. Table~\ref{tab:evaluation_cost_stats} shows the distribution of evaluation times for all three models. On average, \textcolor{linkcolor}{\texttt{gwModel\_kick\_prec\_flow}} requires $\sim0.05$ seconds per evaluation, whereas \textcolor{linkcolor}{\texttt{HLZ}} takes only $\sim0.0025$ seconds, and \textcolor{linkcolor}{\texttt{NRSur7dq4Remnant}} takes $\sim3$ seconds. Thus, \textcolor{linkcolor}{\texttt{gwModel\_kick\_prec\_flow}} (\textcolor{linkcolor}{\texttt{HLZ}}) is approximately 60 (1200) times faster than \textcolor{linkcolor}{\texttt{NRSur7dq4Remnant}}.  

We further find that increasing the number of spin-angle samples causes the evaluation cost of \textcolor{linkcolor}{\texttt{NRSur7dq4Remnant}} to grow linearly, as each configuration is computed sequentially. In contrast, the evaluation times for \textcolor{linkcolor}{\texttt{gwModel\_kick\_prec\_flow}} and \textcolor{linkcolor}{\texttt{HLZ}} remain nearly constant regardless of sample size. Note also that the evaluation cost of GPR model scales as $\mathcal{O}(N_\mathrm{train})$, where $N_\mathrm{train}$ is the number of training data samples, while the evaluation cost of \textcolor{linkcolor}{\texttt{gwModel\_kick\_prec\_flow}} and \textcolor{linkcolor}{\texttt{HLZ}} is indepdendent of $N_\mathrm{train}$.

Our analytic aligned-spin model \textcolor{linkcolor}{\texttt{gwModel\_kick\_q200}} achieves comparable efficiency to \textcolor{linkcolor}{\texttt{HLZ}}, while the GPR-based \textcolor{linkcolor}{\texttt{gwModel\_kick\_q200\_GPR}} incurs similar computational costs to \textcolor{linkcolor}{\texttt{NRSur7dq4Remnant}}. For both computational and extrapolation-related reasons, we do not recommend using \textcolor{linkcolor}{\texttt{gwModel\_kick\_q200\_GPR}} for astrophysical applications.

\begin{figure}
\includegraphics[width=\columnwidth]{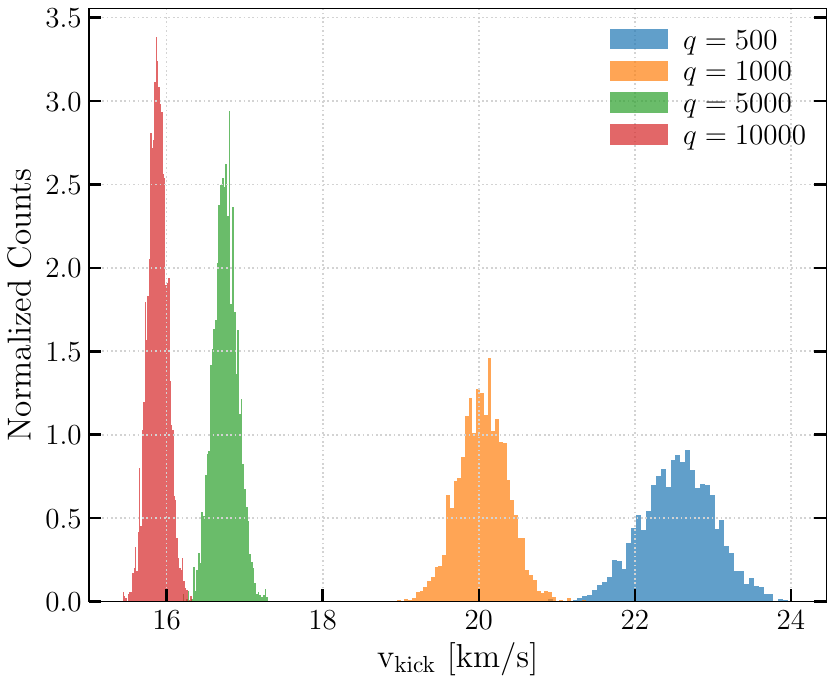}
\caption{Unlike the GPR based models, we find that the \textcolor{linkcolor}{\texttt{gwModel\_kick\_prec\_flow}} model shows reasonable extrapolation outside its training parameter region ($q \leq 100$). Its kick velocity distribution (due to different spin orientation angles) predictions vary smoothly across different mass ratios in the extreme–mass-ratio regime, extending well beyond its training range. For this test, the spin magnitudes are fixed at $|\chi_1| = 0.8$ and $|\chi_2| = 0.9$. More details are in Section~\ref{sec:extrapolation}.}
\label{fig:extrapolation_prec}
\end{figure}

\subsection{Extrapolation capabilities}
\label{sec:extrapolation}
One of the key challenges with data-driven models is their tendency to perform poorly outside their training domain (often exhibiting divergent behavior) or in regions of parameter space with sparse data coverage. This issue becomes more pronounced in high-dimensional spaces such as for precessing binaries where the system is typically described by seven intrinsic parameters. Although our normalizing-flow model marginalizes over the spin-orientation angles, the underlying data still occupy this seven-dimensional space. In previous sections, we demonstrated that GPR–based models such as \textcolor{linkcolor}{\texttt{gwModel\_kick\_q200}}, \textcolor{linkcolor}{\texttt{NRSur7dq4Remnant}}, and \textcolor{linkcolor}{\texttt{NRSur3dq8Remnant}} often fail when extrapolated or evaluated in data-scarce regions of the parameter space. As shown in Figure~\ref{fig:nospin} and Figure~\ref{fig:smooth_aligned_spin_kick}, these GPR models tend to deviate rapidly from the data beyond their training range, producing either unphysical blow-ups or non-smooth behavior.

To assess whether similar issues arise for our normalizing-flow model \textcolor{linkcolor}{\texttt{gwModel\_kick\_prec\_flow}}, we perform extensive tests across a wide range of parameter-space points, including extreme extrapolations up to mass ratios as large as $q = 10^4$, even though the model was trained only up to $q = 100$ (Figure~\ref{fig:extrapolation_prec}). We find that our model’s predictions remain smooth and non-divergent, decreasing monotonically with increasing mass ratio as expected. One could further improve the flow model results to mimic the asymptotic $\eta^2$ scaling expected from analytic studies, however the current extrapolation kick values are sufficient for most astrophysical simulations as they are lower than the typical escape velocities in astrophysical systems. We emphasize that, to our knowledge, this is the first data-driven model in GW astrophysics that exhibits such stable and physically sensible extrapolation behavior far beyond its training domain.

\begin{figure}
\includegraphics[width=\columnwidth]{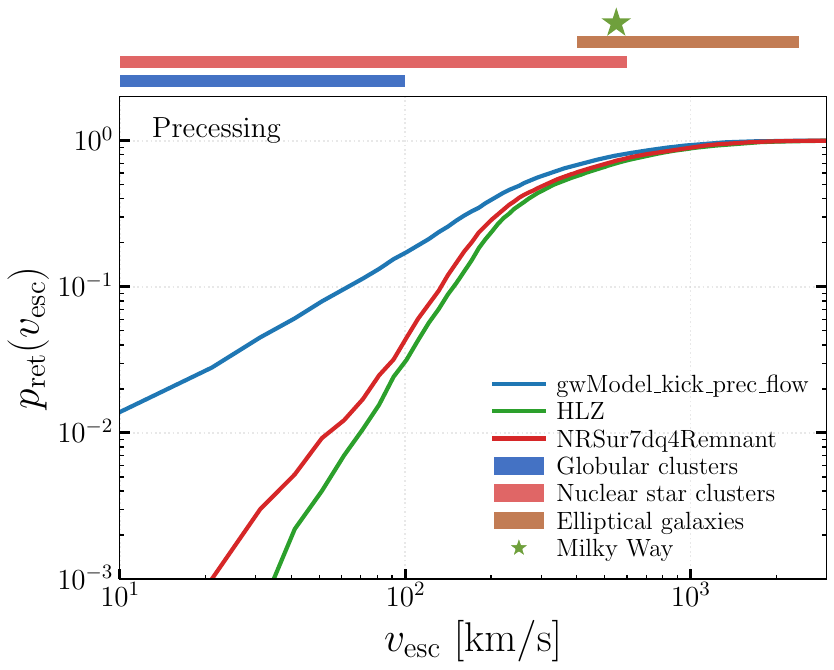}
\includegraphics[width=\columnwidth]{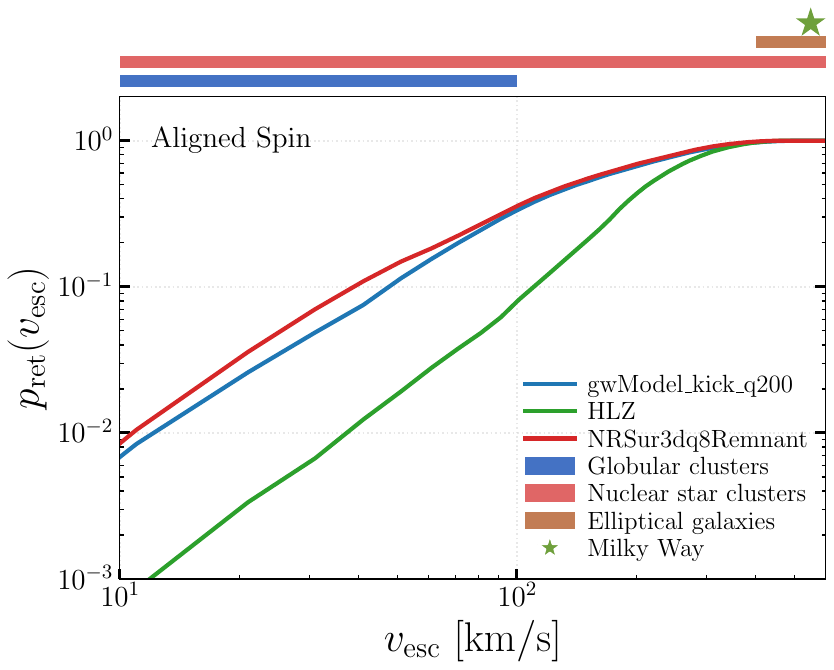}
\caption{Retention probability of first-generation BBH mergers as a function of the host environment's escape velocity. The upper panel shows results for precessing BBHs with kick velocities computed using the \textcolor{linkcolor}{\texttt{gwModel\_kick\_prec\_flow}} (blue), \textcolor{linkcolor}{\texttt{NRSur7dq4Remnant}} (red), and \textcolor{linkcolor}{\texttt{HLZ}} (green) models; the lower panel shows aligned-spin BBHs with kick velocities computed using the \textcolor{linkcolor}{\texttt{gwModel\_kick\_q200}} (blue), \textcolor{linkcolor}{\texttt{NRSur3dq8Remnant}} (orange), and \textcolor{linkcolor}{\texttt{HLZ}} (green) models. Typical escape-velocity ranges for globular clusters, nuclear star clusters, elliptical galaxies, and active galactic nuclei are indicated for reference. All results are obtained using the \textcolor{linkcolor}{\texttt{gwGenealogy}} package. More details are in Section~\ref{sec:back-of-the-envelop}.}
\label{fig:prec_implications}
\end{figure}

\section{Astrophysical Implications}
\label{sec:implication}
We now explore one of the astrophysical implications of our models in the context of hierarchical mergers in dense astrophysical environments such as globular clusters, nuclear star clusters, and elliptical galaxies. We perform both back-of-the-envelope calculations and detailed semi-analytic $N$-body cluster simulations.

\begin{figure}
\includegraphics[width=\columnwidth]{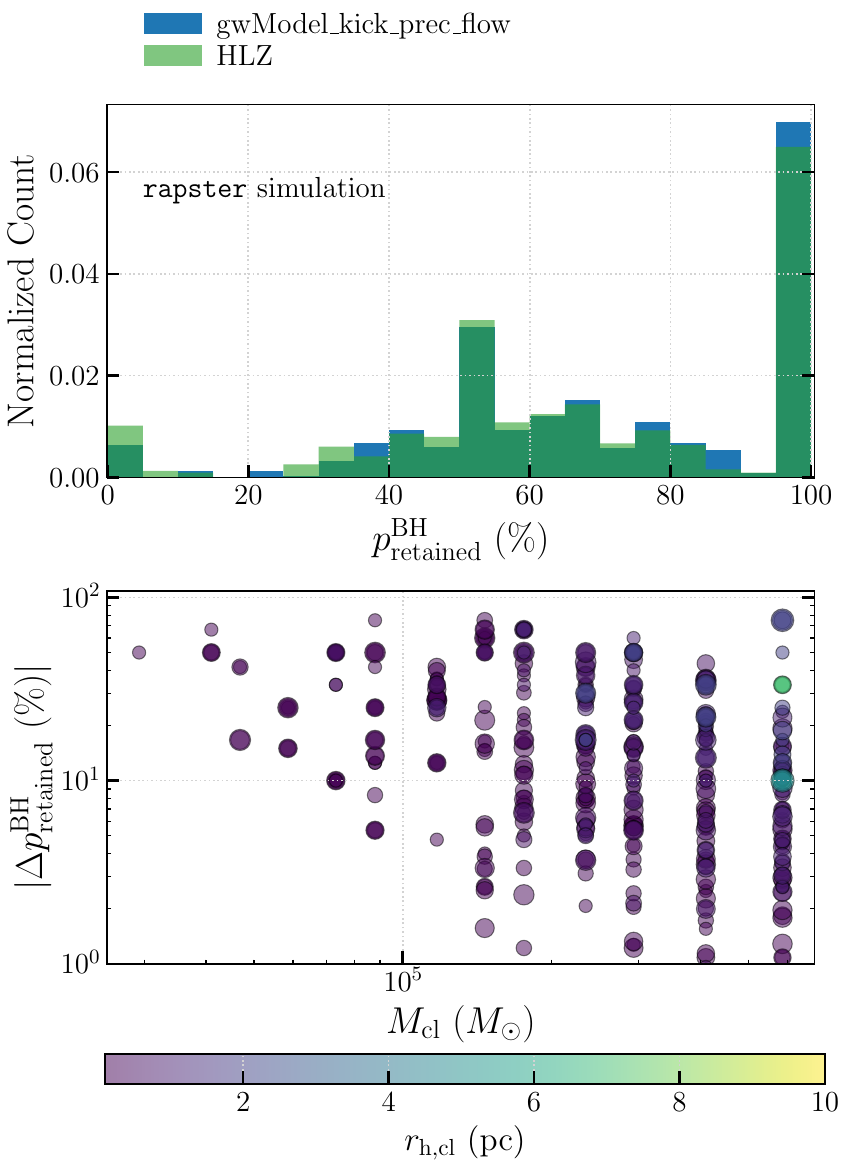}
\caption{\textit{(Upper panel)}: Distribution of BH retention probabilities obtained from detailed stellar evolution and BBH assembly simulations for 1404 different cluster models with varying initial number of stars, half-mass radius, metallicity, and natal spin of 1g BHs. We compare results using two kick-velocity models: \textcolor{linkcolor}{\texttt{gwModel\_kick\_prec\_flow}} (blue) and \textcolor{linkcolor}{\texttt{HLZ}} (green).  
\textit{(Lower panel)}: Distributions of the differences in overall BH retention probability as a function of the initial cluster mass $M_{\rm cl}$, computed using the same two kick-velocity models. Each marker is color-coded by the cluster half-mass radius $r_{\rm h}$, and the symbol size scales with the natal black hole spin $s_{\rm BH}$—smaller circles correspond to lower spins, while larger circles correspond to higher spins. Results are obtained using the \textcolor{linkcolor}{\texttt{rapster}} package. More details are in Section~\ref{sec:nbody}.}
\label{fig:rapster_result}
\end{figure}

\subsection{Back-of-the-envelope calculations for clusters}
\label{sec:back-of-the-envelop}
We first focus on the use of the precessing-spin kick model in the context of binaries in environments such as globular clusters, nuclear star clusters, and elliptical galaxies where binaries are mostly expected to be precessing. For this, we use \textcolor{linkcolor}{\texttt{gwGenealogy}} package\footnote{\href{https://github.com/tousifislam/gwGenealogy}{https://github.com/tousifislam/gwGenealogy}}.
We generate an ensemble of 5000 BHs whose component masses $m_{1,2}$ follow a power-law distribution, $p(m_{1,2}) \propto m^{-\beta_m}$ (with $\beta_m=-2$)~\cite{LIGOScientific:2020kqk}, within the range $[5,\,50]\,M_{\odot}$—representing a fiducial first-generation black hole population formed through stellar collapse. We assign spin magnitudes such that 50\% of the BHs follow a Beta distribution, $p(|\chi_{1,2}|) \propto x^{\alpha-1}(1-x)^{\beta-1}$ (with $\alpha=1.4$ and $\beta=3.6$)~\cite{KAGRA:2021duu}, while the remaining 50\% have spins drawn uniformly from $\mathcal{U}(0,1)$. 
We use a mixed population to ensure that our results are not biased.
The spin orientation angles are chosen isotropically: the polar angles $\theta_{1,2}$ are sampled uniformly in $\cos(\theta_{1,2})$, and the azimuthal angles $\phi_{1,2}$ are drawn uniformly from $[0,\,2\pi)$. We then pair and merge these BHs randomly without imposing any ordering constraints. For each merger, we compute the recoil (kick) velocity using three precessing-spin models—\textcolor{linkcolor}{\texttt{gwModel\_kick\_prec\_flow}}, \textcolor{linkcolor}{\texttt{NRSur7dq4Remnant}}, and \textcolor{linkcolor}{\texttt{HLZ}}. Although some mergers have mass ratios exceeding $q=6$, we include them in the \textcolor{linkcolor}{\texttt{NRSur7dq4Remnant}} sample for completeness. Once the kick velocities are computed, we estimate the remnant retention probability by assuming that remnants with kick velocities smaller than the host environment’s escape velocity are retained.

In Figure~\ref{fig:prec_implications}, we show the retention probabilities as a function of the escape velocities. For reference, we also indicate the typical escape-velocity ranges of various astrophysical host environments, including globular clusters (up to $\sim 60~\mathrm{km\,s^{-1}}$)~\cite{Antonini:2016gqe}, nuclear star clusters (up to $\sim 600~\mathrm{km\,s^{-1}}$)~\cite{Antonini:2016gqe}, and elliptical galaxies (ranging from $\sim 400$ to $\sim 2400~\mathrm{km\,s^{-1}}$~\cite{Merritt:2004xa}). We also mark the approximate escape velocity of the Milky Way ($\sim 500~\mathrm{km\,s^{-1}}$~\cite{Monari2018}). We find that the retention probability of merger remnants increases when using the \textcolor{linkcolor}{\texttt{gwModel\_kick\_prec\_flow}} model, while the \textcolor{linkcolor}{\texttt{HLZ}} model yields the lowest retention probabilities. This trend is consistent with our findings in Fig.~\ref{fig:gwModel_prec_flow_distribution} and Section~\ref{sec:precessing}, where \textcolor{linkcolor}{\texttt{HLZ}} consistently predicted larger kick velocities. Similar results have also been reported in Ref.~\cite{Borchers:2025sid}. Overall, we find that using \textcolor{linkcolor}{\texttt{gwModel\_kick\_prec\_flow}} can increase the remnant retention probability by approximately $5$–$10\%$ for many binaries, which could significantly influence hierarchical merger rates in low-mass globular clusters.

Next, we focus on aligned-spin BBHs in AGNs. Several formation scenarios suggest that, due to interactions with the surrounding gas disk, the component spins in AGNs are more likely to align with the orbital angular momentum axis, leading predominantly to aligned-spin BBHs~\cite{Yang:2019cbr}. We adopt the same mass distribution as in the previous experiment. The spin components along the $z$-axis are now sampled uniformly between $-1$ and $1$. We then generate an ensemble of $5000$ BBHs and merge them randomly. 
For each merger, we compute the remnant kick velocity using three aligned-spin models: \textcolor{linkcolor}{\texttt{gwModel\_kick\_q200}}, \textcolor{linkcolor}{\texttt{NRSur3dq8Remnant}}, and \textcolor{linkcolor}{\texttt{HLZ}}. For reference, we also indicate the typical escape velocities of various host environments, including globular clusters, nuclear star clusters, and particularly active galactic nuclei (between $200$ km/s to $400$ km/s~\cite{Manzano_King_2019}). We find that the \textcolor{linkcolor}{\texttt{HLZ}} model consistently predicts larger kicks, resulting in lower remnant retention probabilities. When comparing \textcolor{linkcolor}{\texttt{gwModel\_kick\_q200}} with \textcolor{linkcolor}{\texttt{NRSur3dq8Remnant}}, we find very similar retention probabilities, with the latter yielding slightly higher values. However, as demonstrated in Section~\ref{sec:aligned_spin}, \textcolor{linkcolor}{\texttt{gwModel\_kick\_q200}} provides more accurate predictions than \textcolor{linkcolor}{\texttt{NRSur3dq8Remnant}} in the aligned-spin regime.

While these back-of-the-envelope calculations provide some insight, they neglect many astrophysical effects that are present in star clusters and AGNs. It is therefore important to perform detailed cluster-evolution modeling that incorporates these effects. We do so in the next section.

\subsection{Semi-analytic cluster simulations}
\label{sec:nbody}
Next, we use the rapid cluster evolution code \textcolor{linkcolor}{\texttt{rapster}}\footnote{\url{https://github.com/Kkritos/Rapster}}~\cite{Kritos:2022ggc} to simulate the evolution of massive star clusters and the assembly of BBHs within them, using more realistic prescriptions than those employed in Section~\ref{sec:back-of-the-envelop}. Unlike detailed $N$-body simulation codes, which are computationally expensive, \textcolor{linkcolor}{\texttt{rapster}} adopts efficient semi-analytic prescriptions calibrated against full $N$-body results, allowing rapid yet physically motivated modeling of cluster evolution.  

We simulate a total of 1404 clusters, varying the half-mass radius $r_h = [0.1,\,0.2,\,0.5,\,0.7,\,1,\,2,\,5,\,7,\,10]$~pc, metallicity $Z = [0.0002,\,0.001,\,0.02]$, and natal spin of first-generation (1g) BHs $s = [0.0,\,0.2,\,0.4,\,0.6,\,0.8]$. 
Furthermore, we choose 13 different values for the initial number of stars, $N_{\mathrm{star}}$, within the range $[5,100]\times10^4$. This corresponds to an initial cluster mass $M_{\rm cl}$ in the range $\sim3\times10^4\,M_{\odot}$ to $\sim6\times10^6\,M_{\odot}$.
The redshift of cluster formation is set to $z = 3$~\footnote{Redshift $z\approx3$ is especially important for star formation because it lies near the epoch when the universe was forming stars at (or close to) its maximum rate~\cite{MadauDickinson2014}.}. The initial stellar population follows a zero-age main sequence (ZAMS) mass distribution $M_{\mathrm{ZAMS}} \in [0.08,\,150]\,M_{\odot}$, and the initial binary fraction is set to $0.1$. Furthermore, for the 1g BHs, \textcolor{linkcolor}{\texttt{rapster}} assigns the spin orientation angles isotropically.
By default, \textcolor{linkcolor}{\texttt{rapster}} computes remnant kick velocities using the \textcolor{linkcolor}{\texttt{HLZ}} prescription. We have now modified the code to incorporate our \textcolor{linkcolor}{\texttt{gwModel\_kick\_prec\_flow}} model for kick-velocity calculations. In \textcolor{linkcolor}{\texttt{rapster}}, the escape velocity of each cluster is determined self-consistently using the virial theorem.

Out of these $1404$ clusters, only $628$ produced BBH mergers.
In Figure~\ref{fig:rapster_result} (upper panel), we show the distribution of BH retention probabilities across these $628$ simulated cluster models when using two different kick-velocity prescriptions. We find that for many clusters, the retention probability can vary significantly depending on the chosen model. The median BH retention probabilities are $54.5\%$ for \textcolor{linkcolor}{\texttt{HLZ}} and $52.7\%$ for \textcolor{linkcolor}{\texttt{gwModel\_kick\_prec\_flow}}, indicating only a modest overall difference. However, a larger fraction of clusters exhibit retention probabilities exceeding $80\%$ when \textcolor{linkcolor}{\texttt{gwModel\_kick\_prec\_flow}} is used. 

Overall, we find that the retention probabilities can differ by as much as $70\%$ across clusters (Figure~\ref{fig:rapster_result}, lower panel). To investigate whether any specific cluster property drives this difference, we plot the variation in retention probabilities as a function of the initial cluster mass and half-mass radius, where the size of each marker (Figure~\ref{fig:rapster_result}, lower panel) indicates the natal black hole spin. We find that in low-mass clusters, fewer black holes are produced, and for these systems, differences between the \textcolor{linkcolor}{\texttt{HLZ}} and \textcolor{linkcolor}{\texttt{gwModel\_kick\_prec\_flow}} kick-velocity distributions almost always affect the retention rates. In contrast, for high-mass clusters ($M_{\rm cl} \geq 10^5\,M_{\odot}$), the differences vary considerably depending on other cluster parameters.

\section{Concluding remarks}
\label{sec:conclusion}
In this paper, we presented two kick-velocity models for aligned-spin binaries: \textcolor{linkcolor}{\texttt{gwModel\_kick\_q200}} and \textcolor{linkcolor}{\texttt{gwModel\_kick\_q200\_GPR}}. The former is an analytic model, while the latter is based on GPR. Both models are trained on a combination of NR and BHPT data up to a mass ratio of $q=200$. We find that both models reproduce the training data accurately, as verified through 5-fold cross-validation, and outperform existing state-of-the-art NR surrogate models \textcolor{linkcolor}{\texttt{NRSur7dq4Remnant}} and \textcolor{linkcolor}{\texttt{NRSur3dq8Remnant}} and analytic \textcolor{linkcolor}{\texttt{HLZ}} model in the aligned-spin limit. Furthermore, our model errors are comparable to the intrinsic NR uncertainties up to $q=128$. Finally, we demonstrate that GPR-based models—including \textcolor{linkcolor}{\texttt{gwModel\_kick\_q200\_GPR}}, \textcolor{linkcolor}{\texttt{NRSur7dq4Remnant}}, and \textcolor{linkcolor}{\texttt{NRSur3dq8Remnant}}—while often yielding low validation errors, can exhibit non-smooth behavior across the parameter space and produce unreliable predictions when extrapolated or in data-sparse regions.

Next, we develop a normalizing-flow–based probabilistic model, \textcolor{linkcolor}{\texttt{gwModel\_kick\_prec\_flow}}, which models the distribution of kick velocities marginalized over all spin orientation angles and conditioned on the binary’s mass ratio and spin magnitudes. Through extensive validation and comparison against the \textcolor{linkcolor}{\texttt{NRSur7dq4Remnant}} and \textcolor{linkcolor}{\texttt{HLZ}} models, we demonstrate that \textcolor{linkcolor}{\texttt{gwModel\_kick\_prec\_flow}} closely matches \textcolor{linkcolor}{\texttt{NRSur7dq4Remnant}} across most of the comparable-mass regime ($q \leq 4$), where the latter is valid. However, \textcolor{linkcolor}{\texttt{gwModel\_kick\_prec\_flow}} predicts noticeably different kick-velocity distributions compared to the \textcolor{linkcolor}{\texttt{HLZ}} model.

Together, for the first time, we provide kick-velocity models for both aligned-spin and precessing binary black holes that remain accurate across the full mass-ratio regime and outperform all existing models. These models enable astrophysical population synthesis and hierarchical merger studies over a wide range of mass ratios without the need to combine models with different validity domains, thereby minimizing potential systematic biases. One current limitation is that our precessing-spin model is probabilistic and marginalizes over spin-orientation angles, as our primary goal here is to support population synthesis and hierarchical merger applications. In future work, we plan to develop an extended version capable of providing point-wise kick predictions for specific precessing-spin configurations. For wider community use, all models presented in this work are publicly available through the \textcolor{linkcolor}{\texttt{gwModels}} package.

We have also performed simplified calculations considering $\sim 5000$ BBHs in clusters and AGNs, computing their retention probabilities as a function of the host escape velocity using our models \textcolor{linkcolor}{\texttt{gwModel\_kick\_prec\_flow}} (for precessing-spin BBHs) and \textcolor{linkcolor}{\texttt{gwModel\_kick\_q200}} (for aligned-spin BBHs). We find that when \textcolor{linkcolor}{\texttt{gwModel\_kick\_prec\_flow}} is used, the retention probability increases by up to $10\%$ across several escape-velocity ranges. For aligned-spin systems, the retention probability is slightly lower compared to \textcolor{linkcolor}{\texttt{NRSur3dq8Remnant}}. In both cases, the \textcolor{linkcolor}{\texttt{HLZ}} model yields significantly smaller retention probabilities. We further performed detailed semi-analytic stellar evolution and BBH assembly simulations using the \textcolor{linkcolor}{\texttt{rapster}} code, modeling 1404 different clusters with varying initial number of stars, metallicity, BH natal spin, and half-mass radius. We again find that the retention probability can change substantially when using \textcolor{linkcolor}{\texttt{gwModel\_kick\_prec\_flow}} instead of the default \textcolor{linkcolor}{\texttt{HLZ}} prescription. Using \textcolor{linkcolor}{\texttt{gwModel\_kick\_prec\_flow}} model in cluster simulations can therefore change hierarchical merger scenarios.

In the future, we plan to extend this work in several directions. First, we aim to incorporate our kick models into different standard star-cluster and population-synthesis frameworks (apart from \textcolor{linkcolor}{\texttt{rapster}}) to study hierarchical mergers in a more systematic and astrophysically consistent manner. Second, we plan to include orbital eccentricity in our kick-velocity models. Although previous studies have suggested that residual eccentricity can enhance recoil velocities based on post-Newtonian and perturbative approximations~\cite{Sperhake:2019wwo,Sopuerta:2006et,Sopuerta:2006wj}, the limited availability of spinning, eccentric NR simulations currently prevents the construction of accurate data-driven models in this regime. We further aim to distill our flow model into a fully analytic model, which will reduce evaluation costs and facilitate both interpretation and implementability.

\begin{acknowledgments}
We thank Gaurav Khanna for generously providing us the BHPT simulations and Konstantinos Kritos for help with population synthesis codes. We are grateful to the SXS collaboration and RIT NR group for maintaining publicly available catalog of NR simulations which has been used in this study. We thank Shadab Alam and Mahajan Rutvik Ashish for pointing out issues with existing kick models and for motivating this work. We thank Maya Fisbach, Carl Rodriguez, Scott Field, and Vijay Varma for helpful discussions.
T.I. is supported in part by the National Science Foundation under Grant No. NSF PHY-2309135 and the Gordon and Betty Moore Foundation Grant No. GBMF7392. 
Use was made of computational facilities purchased with funds from the National Science Foundation (CNS-1725797) and administered by the Center for Scientific Computing (CSC). The CSC is supported by the California NanoSystems Institute and the Materials Research Science and Engineering Center (MRSEC; NSF DMR 2308708) at UC Santa Barbara. D.W. is supported by NSF Grants No.~AST-2307146, No.~PHY-2513337, No.~PHY-090003, and No.~PHY-20043, by NASA Grant No.~21-ATP21-0010, by John Templeton Foundation Grant No.~62840, by the Simons Foundation [MPS-SIP-00001698, E.B.], by the Simons Foundation International [SFI-MPS-BH-00012593-02], and by Italian Ministry of Foreign Affairs and International Cooperation Grant No.~PGR01167.
This work was carried out at the Advanced Research Computing at Hopkins (ARCH) core facility (\url{https://www.arch.jhu.edu/}), which is supported by the NSF Grant No. OAC-1920103. 
\end{acknowledgments}

\bibliography{kick_references}

\appendix

\subsection{Simple analytic expression}
\label{sec:analytic}
As \textcolor{linkcolor}{\texttt{gwModel\_kick\_prec\_flow}} is a data-driven model, we also provide a simple analytic expression, which at this stage is significantly less accurate than 
\textcolor{linkcolor}{\texttt{gwModel\_kick\_prec\_flow}}. The current version of the model is trained on a set of 80 BAM NR~\cite{Hamilton:2023qkv} simulations and 62 SXS NR simulations for single-precessing binaries, i.e., configurations with $\phi_{1} = \phi_{2} = \theta_{2} = 0$. Following Refs.~\cite{Lousto:2008dn,Lousto:2010xk,Lousto:2012gt,Lousto:2012su,Gonzalez:2007hi}, we decompose the total kick velocity as
\begin{equation}
V_{\mathrm{kick,1prec}}(q, |\chi_{1}|, |\chi_{2}|, \theta_{1}) =
\sqrt{V_{\mathrm{kick,AS}}^{2} + V_{\mathrm{kick,\parallel}}^{2}},
\end{equation}
where $V_{\mathrm{kick,\parallel}}$ denotes the component of the recoil due to spin precession.  
We find that $V_{\mathrm{kick,\parallel}}$ can be reasonably approximated as:
\begin{equation}
\begin{aligned}
V_{\parallel} =\;& 
\eta^{2}|\chi_{1}|\left(A_{1}\sin\theta_{1} - B_{1}\sin(2\theta_{1})\right)\\
&+ \eta^{2}|\chi_{1}|^{2}\left(A_{2}\sin\theta_{1} + B_{2}\sin(2\theta_{1})\right) \\
&+ \eta^{3}|\chi_{1}|\left(A_{3}\sin\theta_{1} + B_{3}\sin(2\theta_{1})\right)\\
&+ \eta^{3}|\chi_{1}|^{2}\left(A_{4}\sin\theta_{1} + B_{4}\sin(2\theta_{1})\right) \\
&+ \eta^{4}|\chi_{1}|\left(A_{5}\sin\theta_{1} + B_{5}\sin(2\theta_{1})\right)\\
&+ \eta^{4}|\chi_{1}|^{2}\left(A_{6}\sin\theta_{1} + b_{6}\sin(2\theta_{1})\right).
\end{aligned}
\label{eq:vparallel}
\end{equation}
The best-fit coefficients for Eq.~(\ref{eq:vparallel}), obtained using the \textcolor{linkcolor}{\texttt{scipy.curve\_fit}}~\cite{Virtanen:2020scipy} routine, are
$A_{1} = (3.772 \pm 150.262)\times10^{4}$,
$B_{1} = (1.219 \pm 1.354)\times10^{5}$,
$A_{2} = (3.021 \pm 274.358)\times10^{4}$,
$B_{2} = (2.853 \pm 234.152)\times10^{5}$,
$A_{3} = (-1.594 \pm 18.922)\times10^{5}$,
$B_{3} = (1.177 \pm 15.907)\times10^{6}$,
$A_{4} = (1.154 \pm 33.723)\times10^{5}$,
$B_{4} = (-2.894 \pm 26.947)\times10^{6}$,
$A_{5} = (3.835 \pm 50.846)\times10^{5}$,
$B_{5} = (-3.012 \pm 42.894)\times10^{6}$,
$A_{6} = (-1.110 \pm 91.642)\times10^{6}$,
and $b_{6} = (7.595 \pm 71.445)\times10^{6}$. We denote this model as \textcolor{linkcolor}{\texttt{gwModel\_kick\_singleprec}}.

\begin{figure}
\includegraphics[width=\columnwidth]{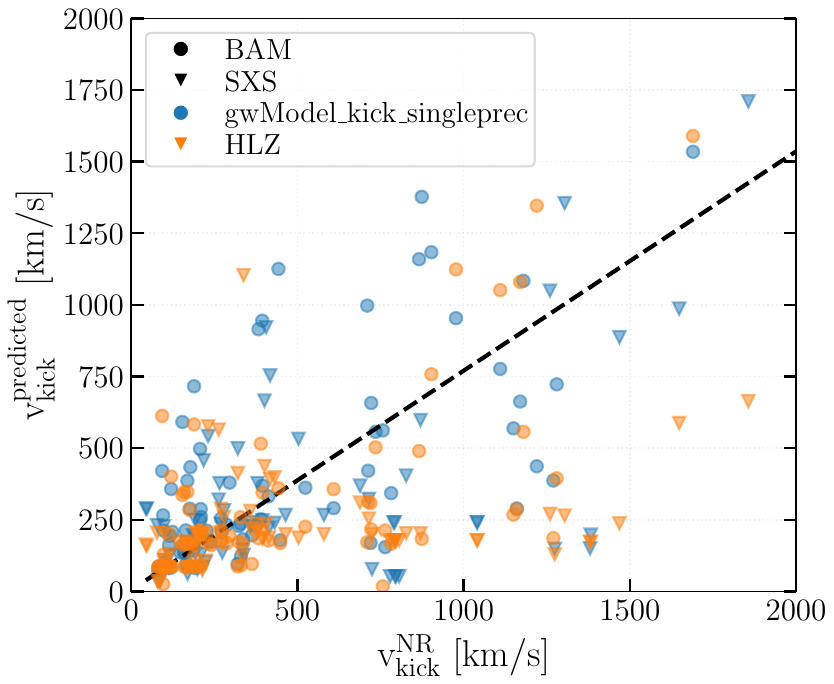}
\caption{We show the true and predicted kick velocities for a set of BAM (circles) and SXS (triangles) NR simulations of single-precessing binaries. The predictions are obtained from the simple analytic \textcolor{linkcolor}{\texttt{gwModel\_kick\_singleprec}} model (blue) and \textcolor{linkcolor}{\texttt{HLZ}} (orange). The black dashed line denotes the line of equality.}
\label{fig:analytic_single_precession}
\end{figure}

In Figure~\ref{fig:analytic_single_precession}, we show the predictions from \textcolor{linkcolor}{\texttt{gwModel\_kick\_singleprec}} alongside the corresponding BAM and SXS NR data. As mentioned earlier, this model successfully captures the overall trend but is not as accurate as the data-driven model developed in this work. Furthermore, we find similar level of disagreement between \textcolor{linkcolor}{\texttt{HLZ}} model and NR data. It is also important to note that, for the NR simulations, we use the initial binary configurations as inputs to our model. In principle, a more consistent treatment of the spin angle $\theta_{1}$ (defined in an appropriate reference frame) should be employed. We leave this refinement for future work. 

\end{document}